\documentclass[10pt,journal,compsoc, hyphens]{IEEEtran}

\ifCLASSOPTIONcompsoc
  \usepackage[nocompress]{cite}
\else
  \usepackage{cite}
\fi
  \usepackage{verbatim}

\usepackage{amsmath}
\usepackage{amsfonts}

\ifCLASSINFOpdf
\else
\fi
\usepackage{url}
\usepackage{balance}
\balance

\usepackage{graphicx}
\usepackage{tabularx}
\usepackage{array}  
\usepackage{enumitem}
\usepackage{hhline}
\usepackage{xcolor}
\usepackage{hyperref}

\newcolumntype{L}[1]{>{\raggedright\let\newline\\\arraybackslash\hspace{0pt}}p{#1}}

\begin{document}
%
\title{Firmware over-the-air programming techniques for IoT networks - A survey}
%
%
%
%

\author{Konstantinos~Arakadakis, Pavlos~Charalampidis, Antonis~Makrogiannakis, Alexandros~Fragkiadakis

%
%

\IEEEcompsocitemizethanks{
\IEEEcompsocthanksitem K. Arakadakis, P. Charalampidis, A. Makrogiannakis and A. Fragkiadakis are with the
Institute of Computer Science, Foundation for Research and
Technology-Hellas (FORTH-ICS), Heraklion, Crete GR-70013, Greece.
K. Arakadakis is also with the Department of Computer Science, University of Crete, Heraklion, Crete GR-70013, Greece.\protect\linebreak
Email: konarak@csd.uoc.gr; pcharala@ics.forth.gr; makrog@ics.forth.gr; alfrag@ics.forth.gr.\protect\linebreak
\IEEEcompsocthanksitem Corresponding Author: Alexandros Fragkiadakis\protect\linebreak
}
}



\IEEEtitleabstractindextext{%
\begin{abstract}
The devices forming the Internet-of-Things (IoT) networks need to be re-programmed over-the-air, so that new features are added, software bugs or security vulnerabilities are resolved and their applications can be re-purposed. The limitations of IoT devices, such as installation in locations with limited physical access, resource-constraint nature, large scale and high heterogeneity, should be taken into consideration for designing an efficient and reliable pipeline for over-the-air programming (OTAP). In this work, we  present  a survey  of  OTAP techniques, which can  be  applied  to  IoT  networks. We  highlight the  main challenges  and  limitations of OTAP for IoT devices and  analyse  the  essential  steps  of  firmware  update  process, along  with  different  approaches  and  techniques  that implement them. In addition, we discuss schemes that focus on securing the OTAP process. Finally,  we  present  a  collection  of state-of-the art open-source and commercial platforms that integrate secure and reliable OTAP.
\end{abstract}

\begin{IEEEkeywords}
Internet-of-Things (IoT), over-the-air-programming, firmware update, code dissemination, delta scripts, firmware image similarity, security.
\end{IEEEkeywords}}

\maketitle

\IEEEdisplaynontitleabstractindextext

%
\IEEEpeerreviewmaketitle

\section{Introduction}
\label{sec:introduction}

The Internet-of-Things (IoT) presents itself as an emerging technology, which is able to interconnect a massive number of heterogeneous smart devices for supporting complex data-driven applications in a variety of domains, such as smart-cities, healthcare, industrial automation, etc. The advances in micro-electromechanical systems that enabled the development of cheap sensors, the progress of wireless communications in the field of Wireless Sensor Networks (WSNs), as well as the growing market demand for machine-to-machine interaction have been identified as some of the key factors that led to the remarkable popularity and adoption of the IoT technology~\cite{Lin_17}.

It is common for the devices forming an IoT network to operate unattended for long periods in variable environmental conditions. Irrespective of the care taken during the development phase, the IoT devices need frequently to be re-programmed over-the-air (OTA), either for resolving bugs or security vulnerabilities (identified after deployment) or supporting different features and/or applications. Failing to do so may result in decreased network performance, security breaches that could compromise the privacy and safety of users, and in general undermine the long-term sustainability of the IoT deployment. However, the dynamic, heterogeneous and resource-constrained nature of IoT networks should be taken carefully into consideration for achieving a dependable and efficient OTA programming.

A fundamental characteristic of the IoT networks is the dynamic changes of the network topology that can happen either because of the energy depletion  of nodes or their inability to communicate with adjacent (neighboring) nodes. Furthermore, the IoT devices that are used in such applications are equipped with scarce resources, such as limited memory, storage and processing power to keep the cost and the battery consumption low. These characteristics impose additional challenges on both the firmware design the nodes run, as well as their update during the lifetime of the network.
 
Regardless of the attention given during the development period, software bugs can occur at any level of the system and stage of the development cycle. As stated by in \cite{dunkels_run-time_2006}, an unexpected combination of inputs that are received by the nodes of a network, can stimulate untested firmware branches resulting in unresponsive nodes that may degrade Quality-of-Service (QoS) or even the integrity of the network. Hence, firmware updates are often released to fix such bugs and security breaches, introduce new functionality, or even change the purpose of the application, completely. 

The latter type of update is a common practice when the behavior of the device needs to be altered dynamically in par to changes of the environment and the available storage is too restricted to accommodate multiple application images. In the most common scenario however, the updates introduce minimal modifications to the firmware code, changing the implementation of a few functions or reconfiguring the application parameters \cite{shi_survey_2011}.

Traditionally, in order to update the nodes of a WSN (being commonly at the core of an IoT system), maintenance personnel had to be dispatched and access the nodes via a serial port or other hardwired back channel. The problem with this solution is that it is not scalable and requires a vast amount of time, which may be intolerable when the update includes fixes for security breaches and should be installed by the motes as soon as possible. Furthermore, the physical access to the nodes may be impossible sometimes, since they can be located in inaccessible areas (e.g. implanted into asphalt roads), or implanted into human bodies as medical sensors \cite{wilson_sensors_1999}.

Due to the above limitations, in the early 2000's the first wireless update schemes for restricted embedded devices were developed. These schemes required the firmware image to be built at a base station, that would afterwards transmit it in its entirety to the neighboring nodes over a radio channel. Once a node had fully received the update, it rebooted, the bootloader overwrote the contents of the internal flash memory (program memory) with the new image and started running the new firmware. A limitation of these first wireless solutions \cite{crossbow_technology_inc_2003_mote_nodate}, however, is that only the nodes that were within the radio range of the base station (over a single hop) were able to receive the update. Nowadays, due to the large geographical scale of modern IoT networks, multi-hop communication is a reality, and in order to facilitate proper OTAP, new \emph{dissemination protocols} have been designed, whose goal is the reduction of the energy consumption upon an update, avoiding redundant transmissions and collisions that can degrade the quality of the channel.

In addition, as the developers used to update the firmware frequently, the approach of transmitting the entire firmware image each time a new update had been released, quickly became obsolete, as it seriously affected the lifetime of these devices, which were supposed to run even for years. This limitation, accompanied by the vast amount of time the update of a network required \cite{stolikj_efficient_2013}, triggered the development of the first \emph{incremental programming} schemes. These schemes avoid sending the whole firmware image every time a new update has been released and just transmit commands to the nodes, that instruct them how to reconstruct the new firmware locally, utilising parts of the currently run firmware, that each node has already stored in its flash memory.

Thus, in order to update the modern IoT networks incrementally, a base station should first create the new firmware image and then the resulting delta script, computing the common segments between the new and the previous firmware versions. Afterwards, the delta script is disseminated in the network, utilising a multi-hop protocol, in order to reach all the nodes. Whereupon, each node should interpret the received script, execute the commands found inside, and reconstruct the new firmware locally. Once this last step has been completed, the node can be updated replacing the firmware it currently runs, with the one it just reconstructed (loading phase). This process can be visualized in Figure~\ref{fig:update}.

It must also be noted the importance of the security during the update of the IoT devices. Since most of the protocols that have been designed for this environment aim the update image to reach a lot of the nodes of the network (if not all of them), they follow a propagation approach where each node just needs to receive a part of the update and it forwards it to its neighbors. Although this technique aids the fast update of the network, it raises serious concerns regarding the security of the process. If the nodes do not validate the authenticity of the updates they receive, this could result into the installation of faulty or malicious firmware in the nodes, originating either from outsiders or already compromised network devices.

However, when a new update scheme is designed, the limitations that bind these devices~\cite{liu_implementing_2004}, due to their restricted nature, should be taken into account. For example, actuators and sensors are devices with low-power antennas and limited transmission range, able to reach only a portion of the other nodes in the network. Furthermore, these devices are severely constrained in terms of computational power, memory and storage \cite{noauthor_rfc_nodate}. These characteristics of theirs, dramatically restrict the features that can be embedded into the firmware image and the update mechanism itself \cite{baccelli_scripting_2018}. In general, IoT OTAP is not a trivial task and poses many challenges that can affect the quality and sustainability of the network. The main challenges and limitations that complicate the OTAP process and have been reported by the literature are presented in Section~\ref{sec:limitations}.

In this survey, we have examined the main stages of the wireless update process, mainly focusing on the severely restricted IoT devices. For each such stage, we also present major contributions that have been designed for this class of devices, as well as the limitations and complications the researchers of the  had to overcome due to the restrictions of these devices. In this way, one should be able to understand the underlying motives of design decisions of the corresponding scheme, setting the basic principles for designing OTAP solutions.

We advocate that the division of the OTAP process we have followed (found in Figure~\ref{fig:update}) is the most suitable, as it presents the reasonable flow for updating the motes in an incremental fashion. In contrast to this survey, most surveys usually target a specific aspect of the OTAP process, neglecting the other stages and their interconnection. For example, the authors of \cite{qiang_wang_reprogramming_2006}, divided the update process in a similar way to ours, but specifically targeted the dissemination stage of the update, neglecting the delta generation and the security aspects of the process.

Additionally, in \cite{brown_software_2013} the authors have presented an extensive overview of many update dissemination protocols for WSNs, as well as stand-alone update schemes. However, they avoided presenting the internals of various schemes in higher detail and they compared the protocols and schemes using the same metrics. We believe that such a comparison is not appropriate because most contributions introduce novelties, concentrating on different aspects of the update process and they are often supposed to be used as parts of an update scheme. Thus, a more careful division and comparison of them is required. 


On the other hand, in \cite{zandberg_secure_2019} and  \cite{wang_survey_2006}, the security aspects of the update process were discussed, providing valuable information on how an adversary can exploit the epidemic nature of the dissemination protocols to initiate attacks (e.g. Denial-of-Service (DoS), install malicious code etc.). Moreover, the authors have provided information about the available cryptographic libraries that are suitable for constrained devices as well as their memory footprint and performance. However, this survey does not present any actual contributions that have utilised these libraries to ensure the authenticity and the integrity of the transmitted data during the update.


In~\cite{chen_survey_2016}, a lot of security-oriented dissemination protocols are presented along with some authentication and freshness verification methods that have been used by the literature. Finally, the authors in~\cite{Bauwens_20}, focus on the key principles of OTAP in IoT networks; however, limiting their contribution in brief description of these principles, without presenting and analysing specific research contributions in depth.

Although the surveys mentioned above provide valuable information for the insights of the wireless update process in IoT networks, as well as the related literature, our survey differs in several aspects. The main contributions of this survey paper focus on OTAP techniques, proposed over the period 1999-2020, and are as follows:

\begin{enumerate}

	\item  We perform a comprehensive organisation of the OTAP process in the following steps:
    
    \begin{itemize}
        \item Preservation of the similarity of the two firmware images;
        \item Computation of the delta script, using differencing algorithms to detect common segments;
        \item Dissemination of the update to throughout the nodes of the network.
    \end{itemize}

	\item  We highlight the insights and challenges of each step, as well as analyze extensively and compare contributions related to each step.

	\item  We discuss the limitations of the restricted embedded devices that must be taken into account when designing a new update scheme.

	\item  We highlight the security aspects of the update process and present contributions that ensure the integrity and authenticity of the received  firmware binary during the dissemination stage.
	
	\item  We present and compare some state-of-art open-source and commercial IoT platforms that integrate secure and reliable OTAP support.

\end{enumerate}

\section{Main challenges and limitations for over-the-air programming}
\label{sec:limitations}

\subsection{Limited memory, storage and processing power} 

Usually, IoT nodes are restricted embedded devices with limited memory and storage size. Such nodes typically feature a relatively low-cost internal NAND-based flash memory \cite{sanvido_nand_2008}, called \emph{program flash} that accommodates both the bootloader (piece of software responsible for writing the new firmware code in memory) and the firmware code. Moreover, there is an SRAM or DRAM for the storage of the volatile data, including the heap, the stack, and the global variables of the application. These sections are mapped to predetermined RAM regions when the \emph{reset handler} executes~\cite{yiu_introduction_2015}, a purpose-specific code, whose goal is to prepare the RAM and initialise the registers before the actual application code starts running. Very often, the available RAM in IoT devices is too limited to accommodate the firmware code, so the flash memory is used for this purpose, instead.

Furthermore, IoT nodes are often equipped with an external non-volatile EEPROM for the storage of various data, such as routing tables and other network-related data. Moreover, many wireless update schemes use the EEPROMs for the storage of rollback images (golden image) to provide fail-safe updates.

Additionally, EEPROMs are also used by many OTAP schemes~\cite{panta_hermes_2009, panta_zephyr_2009, shafi_no-reboot_2012} as a temporary storage for the update image. In these schemes, when the image has been received completely, the node copies the new firmware code from EEPROM to the program flash and then starts executing the new firmware. This strategy allows nodes to remain operational while receiving the new firmware image. On the other hand, other schemes (e.g.~\cite{frisch_over_2017}) explicitly use the program flash to store the new firmware along with the currently running one (the flash memory is divided into two equally size regions).


Finally, due to their low cost, IoT nodes typically feature micro-controllers that operate at a lower frequency compared to traditional CPUs~\cite{farooq_operating_2011}. Thus, if such a node executes a complex algorithm that needs intensive processing (e.g. RSA~\cite{nisha_rsa_2017},  AES~\cite{dsouza_2017}), the time overhead will be significantly high, during which the node may be unable to perform any other operations. To address this limitation, the research community works towards lightweight algorithms in software (e.g.~\cite{mughal_2018}), as well as hardware-based acceleration of cryptographic operations (e.g.~\cite{Banerjee_2019}).

\subsection{Flash memory degradation} 
Another challenge for the firmware update process is the quality degradation of the flash memory, caused by the large number of erase and write operations during an update. Flash memories consist of erase units called \emph{blocks}. Prior to writing into a specific block, it must be erased; however, the number of times a block can be erased is limited. Typically, NAND-based flash memories offer at max 1 million flash cycles, 10 times the life of a NOR-based memory~\cite{tal_white_2002}. When the erase threshold of a block has been reached, it is marked as a \emph{bad block} and cannot be used in the future, thus limiting the available storage.

In order to prolong the lifetime of flash memory, blocks must be utilised with caution when a new firmware image has to be stored, so one can achieve a uniform and smooth degradation of the available blocks. Some schemes(e.g.~\cite{ondrej_effective_nodate, dong_elon_2010, kulkarni_mnp_2005,stathopoulos_remote_2004}), exclusively use the RAM for storing temporal segments of a firmware image, avoiding redundant access and unnecessary operations on the flash memory.

\subsection{Energy consumption} 
IoT nodes are often battery-operated, thus constrained in terms of energy, while operating unattended in harsh environments for long periods of time~\cite{ersue_management_2015}. Furthermore, they often rely on ambient power sources such as solar energy, wireless energy and RF to operate uninterrupted. A node's lifetime is highly affected by routine operations such as those involved into the wireless radio communication. As an example, the transmission of a single bit of data could consume roughly the same amount of energy as executing 1000 instructions~\cite{reijers_efficient_2003}.

Writing data into the flash memory is not a lightweight process either, as a higher voltage is required; hence, the total power consumption is significantly affected by the size of the data to be stored, and for this reason, firmware size minimisation is of utmost importance. Furthermore, the update process can be energy intensive due to the amount of messages that need to be transmitted in the network, so firmware dissemination protocols must be carefully implemented to avoid redundant transmissions (e.g. flooding).

Moreover, firmware decompression at the receiving node can consume a significant amount of energy; hence, rendering compression-based update strategies unsuitable for use by constrained IoT nodes. Finally, node's energy consumption can be affected by the firmware code layout in the flash memory. In~\cite{pallister_high-level_2014}, the authors show that crossing a page boundary in flash memory causes additional energy consumption due to the extra circuitry powered up to read the new page. This implies that the loops in the firmware code must be properly aligned in a page, if possible, to avoid an excessive energy consumption. 



\subsection{Overhead due to node reboot} 
In several OTAP contributions (e.g.~\cite{dong_r3_2013, dong_r2_2013, panta_hermes_2009, panta_zephyr_2009}), after the installation of a new firmware version has been completed, the node is required to reboot, so as the bootloader is run and the new firmware starts executing. This can introduce a significant overhead in time-critical IoT applications such as those used in aviation, health-care, connected cars, etc. Several other OTAP contributions (e.g.~\cite{zhang_live_2016}) address this issue by proposing mechanisms able to dynamically patch an IoT node with a new firmware while it is operating, without the need to reboot.

\subsection{Group-wise IoT node re-programming in heterogeneous environments} 
Another challenge for the firmware update process is that IoT nodes within a network can execute different firmware versions, have different roles and hence have different update requirements. Moreover, these devices can be heterogeneous in terms of software and hardware~\cite{hahm_operating_2016,farooq_operating_2011}. Under these conditions, the update mechanism must have the means to support the update of individual nodes in case of different firmware versions, and simultaneous updating in case of a group of nodes that have a common version. Group programming can minimise the energy consumption and the total re-programming time required.

However, this is not trivial task as an efficient dissemination algorithm should be able to select optimum routing paths, in order to disseminate a new firmware version to a number of nodes. To make this feasible, modification of routing protocols may be required (e.g.~\cite{aschenbruck_selective_2012}). In some contributions (e.g.~\cite{zhu_detection_2018}), the nodes can take decisions based on criteria like the current firmware version, the number of reachable neighbors, etc., to decide if they will accept the update. 

Most contributions, however, follow a two-stage approach to provide group-wised updates. Initially, metadata-carrying packets are flooded in the network and interested receiving nodes respond back. Following this approach, a routing tree is established and the nodes along this path act as firmware forwarding nodes~\cite{qiang_wang_reprogramming_2006}, and the transmission of the actual firmware then starts. Finally, it must be noted that the update methods in such heterogeneous environments have to be portable across hardware platforms and operating systems nodes are based on, avoiding dependencies on libraries (e.g. cryptographic libraries ~\cite{frimpong_iot-cryptodiet_2020,lakkundi_lightweight_2014}) and underlying protocols.

\subsection{Network flooding} 
IoT nodes often communicate in low bandwidth channels and in frequency bands that are overcrowded (i.e in the unlicensed bands of 2.4 GHz) with significant interference. Moreover, the nodes can be dispersed in large geographical areas, for example, in IoT-enabled smart city applications~\cite{TragosBC_17}. OTAP protocols should avoid flooding the network during the update process, by employing techniques for: (i) efficient selection of the nodes to be updated, (ii) firmware compression, (iii) dynamic patching, (iv) energy-aware routing for path selection, etc.

\subsection{Data and state consistency} 
There are various algorithms (e.g. trust-based schemes~\cite{Fragkiadakis_16}, routing algorithms~\cite{Lili_18}, etc.) where nodes cooperatively report measurements, so a central server can infer about a possible event. During the firmware update process, these applications can be severely disrupted, as some nodes may execute the new firmware version, while others the current one. For this reason, extra care should be taken during the update process and a suitable mechanism has to be used to guarantee data and state consistency. Usually, after an update, nodes reboot, so data and state are reset. Moreover, IoT networks are susceptible to changes of the environmental conditions ~\cite{Szewczyk2004LessonsFA} that can result to temporary node disconnections.

\subsection{Security}
IoT proliferation has significantly extended the attack surface and many attacks with disastrous results have already taken place (e.g. Mirai and other botnets~\cite{Antonakakis_17,angrishi_turning_2017}). Having compromised thousands of IoT devices around the world, an attacker can demonstrate large-scale DDoS attacks against critical infrastructures; hence, it is of paramount importance IoT nodes to be supported by an OTAP mechanism for securing bug fixes. Moreover, the OTAP mechanism itself has to be securely designed, otherwise it can become an additional attack vector for an adversary. For example, the Zigbee Worm~\cite{ronen_iot_2017} was able to trigger a chain reaction of infections, initialised by a single compromised IoT device (light bulb), using a malicious firmware update image.

\section{Over-the-air programming essential operations}

\begin{figure*}[htb]
    \centering
    \includegraphics [width=18cm] {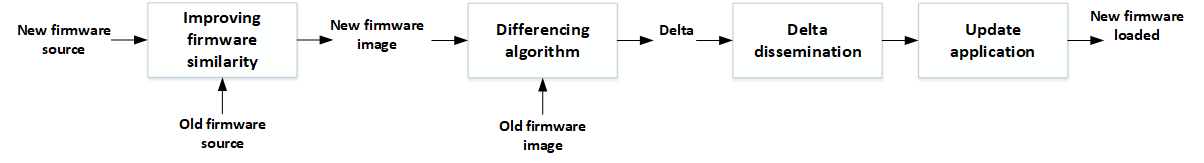}
    
   \caption{Firmware update process essential stages~\cite{ondrej_effective_nodate}}
    \label{fig:update}
\end{figure*}

The limitations imposed by the inherent nature of IoT networks, as described in Section~\ref{sec:limitations}, dictate the careful design of an OTAP pipeline that mainly focuses on firmware image size reduction, as well as efficient and robust dissemination in the network. The essential stages for performing a firmware update, which are illustrated in the form of a flow-graph in Figure~\ref{fig:update}, are~\cite{ondrej_effective_nodate}:
\begin{itemize}
    \item \textbf{Firmware similarity improvement}, which includes comparing the new firmware source code with the old one, in order to mitigate function or variable shifts and increase the similarity between the built firmware versions;
    \item \textbf{Differencing algorithm application}, which includes producing a so-called \emph{delta script} by using a differencing algorithm for comparing the old firmware image with the new one. The delta script encodes a set of instructions that, once applied on the old firmware, enable the reconstruction of the new one. In general, a delta script should be of minimal size (smaller than the original firmware image), since the goal is to reduce the data transmitted to the node;
    \item \textbf{Delta script dissemination}, which is responsible for orchestrating the efficient and reliable firmware (delta script) distribution to the IoT nodes by applying a suitable dissemination protocol that focuses on transmitted data minimisation;
    \item \textbf{Update application}, which refers to the OTAP stage that takes place on the node and includes reconstructing, verifying, installing and executing the new firmware.
\end{itemize}
This section provides a detailed analysis of techniques related to all four stages of the OTAP pipeline and highlights both their advantages and their limitations.

\subsection{Improving firmware images similarity}
Despite the fact that a firmware update can introduce small modifications to the firmware code, these changes can result to a disproportional increase in the size of the delta script produced. As an example, suppose that a software developer modifies a single function within a piece of software that will be executed in an IoT node. As a result, the following instructions will be inserted in the delta script for transmission:  

\begin{itemize}
    \item \textbf{\emph{COPY}} the firmware image segment from address 0x00000000 until the address where the function is located.
    \item \textbf{\emph{ADD}} the segment that contains the new function implementation from the new firmware image.
    \item \textbf{\emph{COPY}} the firmware image segment from the address following the end of the function implementation, until the end of the current firmware.
\end{itemize}
Although the above sequence of operations is correct, this constitutes a poor representation of the resulted delta script because in the binary image of the updated firmware, the code that follows the modified function's implementation will be relocated (shifted), to comply with the new size of the updated function; thus, functions that are located after the modified one, will be shifted to other memory addresses (different from the ones in current version). Subsequently, the instructions that call these functions will use different target addresses (the address of the called function in the flash memory) in the two versions. Finally, since all these modified target addresses will be encoded using \emph{ADD} instructions in the script, the size of the generated delta script will be significantly high. This is also true when a new global variable is defined, or a previously defined one is removed, as other global variables may need to relocate and hence the instructions that reference them will have different target addresses.

In~\cite{shafi_no-reboot_2012}, the authors distinguish four distinct properties of the firmware image that affect the size of the generated delta script.
 
 \begin{itemize}
    \item \textbf{Function shifts}. When a function is modified, either shrunk or grown, other functions located into lower memory addresses are forced to relocate; hence, the target addresses of the corresponding calling instructions need to change. In this case, a \emph{COPY} instruction will be injected into the delta script to encode the common segments present in the two firmware versions, and an \emph{ADD} instruction will be required to encode the bytes associated with the target addresses.
    
    \item \textbf{Global variables shifts}. The insertion of new global variables within the code can also affect the structure of the resulted binary image. Global variables are stored in RAM in the \emph{.data} section when initialised, followed by the \emph{.bss} section, where the uninitialised variables are mapped to. This is a common binary representation format of the different memory sections. When a new global variable is initialised, all other variables located in subsequent RAM locations, will be shifted and the references (by other functions) to them will have different target addresses for the two firmware images. Hence, a declaration of an initialised global variable will effectively shift all variables in the \emph{.bss} section and the ones following within the \emph{.data} section. Moreover, a firmware programmer has no control over the placement of the global variables in RAM, since this is determined by the compiler and not by the order they are declared in the source code~\cite{panta_hermes_2009}.
    
    \item \textbf{Relative jumps (inside functions)}. 
    Inside the source code, the jumps from one function to another are performed using an offset value that represents the distance between the jump instruction and the target address.
    For this reason, inserting new instructions or deleting others between a jump instruction and its target, can result to a different offset, which requires additional \emph{ADD} instructions for the encoding.

    \item \textbf{Indirect addressing} 
    In RISC architectures, memory locations can only be accessed indirectly through the processor registers. For example, if an instruction calls a function, this function's address will be stored in a register and then, the control circuit will enforce a jump based on the value this specific register contains. Since, as stated above, the functions and global variables can be shifted in an update, registers' values need to be updated; thus additional \emph{ADD} operations are required.
\end{itemize}

Based on the above, it is important to preserve images' similarity, mitigating the effects of function and variables shifts, prior to calculating the common segments, in order to create small-sized delta scripts. Various techniques are proposed in the literature for keeping the target addresses unchanged. Some contributions presented later in this section, transmit only the segments that have been modified in the new version  and not the whole update image. In these schemes, when a node receives such a firmware segment, it either re-links the image locally, or patches the image using the received information. Moreover, these updated segments can also be used as input to a differencing algorithm, to further reduce the update overhead. 

Most OTAP contributions avoid direct modifications of the compilers and linkers, as they are provided by hardware vendors, designed to deliver highly optimised code~\cite{kachman_configurable_2016}. Usually, the files produced during the code compilation and linking are used. For example in~\cite{park_non-invasive_2014}, the MAP file along with the binary files produced before and after the static linking are used. A MAP file contains the relative offset and length of each object in the firmware and it can be produced by the linker, providing additional valuable information regarding memory usage~\cite{noauthor_map_nodate}. Furthermore, the binary files produced during the static linking can provide information regarding code re-locations, a useful input for schemes aiming to improve firmware similarity.

Finally, many related works (e.g.~\cite{panta_hermes_2009, panta_zephyr_2009, shafi_no-reboot_2012}) target specific boards or platform families, although claiming platform independence. Nevertheless, platform independence cannot be achieved easily, as different micro-controller families use different relocation types and their definitions should be taken into account by the corresponding scheme. Platform-specific techniques have also been developed (e.g.~\cite{adly_over--air_2010, frisch_over_2017}). For example, the authors in~\cite{parthasarathy_over_2010} describe the effort for porting the Deluge protocol to support the Imote2 sensors~\cite{parthasarathy_over_2010}, which was initially designed for the Tmote sky~\cite{noauthor_tmote_2006} and MicaZ~\cite{noauthor_micaz_nodate} motes.  

In the next sections, we briefly describe techniques used to improve the similarity of firmware images. Moreover, in Table~\ref{tbl:OTAP_deltas} some OTAP schemes are presented, along with the similarity preserving technique and the differencing algorithm they have utilised, as well as some other characteristics.

\subsubsection{Slop regions} 
The use of the \emph{slop regions} was first introduced in~\cite{koshy_remote_2005} and since then, it has been adopted by other OTAP schemes as a way to address function and variables shifts. A slop region is defined as the free memory space, located immediately after a function's code in the flash memory, where a function can grow or shrink without causing any other functions to relocate. If a function grows, part of its slop region will be utilised, thus without causing any other function shifts. If a function shrinks, its slop region will grow, occupying the removed part of the function, so other functions that follow will not be shifted. Besides the \emph{.text} region that is mapped in the flash memory, slop regions have also been used between the \emph{.data} and the \emph{.bss} sections in RAM, to avoid global variable shifts ~\cite{panta_hermes_2009}. Finally, in order to implement this feature, the linker has to be modified, with the risk to downgrade the produced code performance.

A more efficient use of the slop regions is presented by the authors of the Qdiff OTAP scheme~\cite{shafi_no-reboot_2012}. To deal with function shifts, Qdiff does not create slop regions for each function during linking, like other implementations do, but when a function is deleted or shrunk, the resulted available space becomes a slop region. Hence, slop regions can be found only immediately after functions' code or inside functions, as a result of removed instructions. On the other hand, if a function grows, QDiff will try to find a slop region right after the function implementation and if such a regions exists, the function will expand there; otherwise, the new code will be moved at the end of the existing code. Moreover, as an update can create (or expand) some functions, while it can delete (or shrink) some others, QDiff proactively creates empty slop regions that are later assigned to functions.

A drawback for using slop regions is that excessive fragmentation of the memory space can occur, as some regions may contain code, while others remain idle. Apart from the inefficient use of the flash memory, fragmentation can increase the energy consumption because the control circuitry needs to activate a large number of memory regions. The authors in~\cite{kachman_configurable_2016} show that the energy consumption can increase by up to 5\% when memory is fragmented. Finally, extra care is needed when a function grows beyond its slop region and might need to relocate to a completely different memory region.

\subsubsection{Position independent code (PIC)}
\emph{Position independent code} is an option that can be set during code compilation, where code is compiled to execute normally, regardless of the absolute memory address it is stored in. All references and target addresses are related to the memory address of the calling instruction; thus, if shifts occur, the "relative" target addresses are not affected. Nevertheless, due to hardware limitations of the embedded devices, these relative (instruction) jumps can be performed only within certain offsets. For example, Atmel AVR platform supports PIC, but restricts the size of the program to 4 KB. The PIC technique is used by the SOS operating system~\cite{han_sos_2005} in order to avoid the effect of address shifts.  

\subsubsection{Indirection tables}
The \emph{indirection tables} technique was first proposed in~\cite{panta_hermes_2009,panta_zephyr_2009} as a countermeasure against function shifts. During the firmware image linking, an indirection table is created, stored in a fixed location within the flash memory. In this table, there is one entry for each function called at least once, along with the memory address it is stored in. Any calls to the functions are replaced by suitable jumps to the corresponding entries of the table. 

The advantage of this technique, as a similarity preserving mechanism, is that when a function relocates, only its entry in the indirection table is affected (memory address updated), while the calling instructions are not affected. However, this technique is platform-specific and linker modification is required. Moreover, because the function calls are performed indirectly, through the table, the run-time latency of the function call increases. Finally, the table size is proportional to the number of functions called, so in complex programs with many function calls, a large amount of flash memory is occupied (by the table).

\subsubsection{Interrupt service routines pinning}
\emph{Interrupt service routines} are software methods, invoked by hardware, to respond to specific interrupts (e.g. packets received by the network adapter). The memory addresses for these services are stored in an interrupt vector table, which in most embedded systems, is placed at the beginning of the program memory. Whenever an interrupt occurs, the control goes to a pre-defined entry of the vector, and through it, the correct interrupt service routine is invoked. However, modifications in the firmware code can relocate these service routines, affecting the memory addresses contained in the interrupt vector table. The authors in~\cite{panta_zephyr_2009} address this issue by mapping the interrupt service routines to fixed memory locations in the program flash.  

\subsubsection{Global variables' address pinning} 
This technique was proposed in \emph{Hermes}~\cite{panta_hermes_2009}, as a way to ensure that the global variables appear in a specific order and hence stored in the same address in each firmware version. The actual order of the global variables in the RAM memory is determined by the compiler type. This technique exploits the fact that members of a structure are placed in the same order in RAM, exactly as they are declared in this structure. The defined, as well as the undefined global variables, are detected and stored into two distinct structures, so if the update does not define additional global variables, it is ensured that the memory addresses of the current variables are not affected. 

Furthermore, Hermes utilises a slop region between the \emph{.data} and \emph{.bss} sections to avoid  address shifts of the undefined variables when the \emph{.data} section shrinks or expands. In another related contribution (\cite{shafi_no-reboot_2012}), the authors follow a different approach to address data shifts. They modified the method for the \emph{.data} and and \emph{.bss} sections expansion in RAM. Instead of expanding towards the same direction, the two sections expand towards opposite directions. To implement this concept, the two sections are placed into a fixed address (initial address) with a large empty space between them (in RAM); thus, when a new initialised global variable is created, is placed at the bottom of the \emph{.data} section, whereas when an uninitialised one is created, is placed at the beginning of the \emph{.bss} section, so no data shifts occur.



\subsubsection{Relocatable code} 
When building a runnable program, such as a firmware image, various modules must be compiled and linked together to construct the final executable. These modules, referred as \emph{relocatables}, are initially assembled at pseudo-addresses and when linked together, the linker resolves these address references to the correct values. However, once the final program is created (after linking), it should also be able to execute from different memory addresses, as multiple users may wish to run multiple instances of the same program; hence, \emph{relocatable code} is a piece of software  whose execution address can be dynamically moved around the available address space and loaded in multiple addresses. In addition, a relocation table is created, containing all these memory references and is used by the loader to resolve the references to the correct absolute addresses, when the program executes.

In~\cite{dong_r2_2013}, the authors utilise the relocatable code technique to mitigate the effect of function and variable shifts. The key idea is to change all references to symbols (functions and global variables) to the same (predetermined) value and provide the needed metadata, so that the loader at the receiving side, properly resolves the references before code execution. The pipeline the authors follow to create the firmware image consists of a three-stage process: first, the linker produces relocatable code, where the resulted relocation table contains an entry for each reference. Then, the target addresses of all reference instructions are changed to zero. Finally, the relocation table and the altered image are merged to form the final image that will be disseminated. 

In~\cite{dong_r3_2013}, the authors use the relocatable code technique more efficiently, minimising the metadata transmission overhead. The authors observed that many instructions reference the same symbols; so instead of filling these references with zeros, they are filled with the corresponding symbol index, which is a reference to a symbol entry that contains the actual address of the symbol. Moreover, this scheme manages the symbols and ensures that their index will not be altered between different firmware versions. Finally, instead of using the 2-byte offset field to yield which memory location needs relocation, for compression reasons, a bitmap is used to indicate which 2-byte memory locations needs relocation.


A key advantage of the relocatable code technique is that it handles more types of reference instructions, in a more general way, and hence is able to perform better than other similarity preserving techniques. A disadvantage, however, is that it requires a sophisticated loader that can resolve the relocated addresses before firmware executes, and the relocation table introduces additional transmission overhead. Finally, it must be mentioned that although the use of relocatable code and indirection tables seem similar, they have some major differences. The relocatable code must be resolved during load time, while the indirection table-based code operates in run- time, introducing an extra overhead. Furthermore, the metadata of the relocatable code are larger in terms of size compared to those of the indirection table-based code.

\subsubsection{In-place patching} 
The authors in~\cite{zhang_live_2016} propose an OTAP strategy based on \emph{in-place} code updating, using code patches to avoid system reboots and make use of the available memory more efficiently. The key idea is to transmit only the parts of the firmware that have been altered (patches), and the update module that executes on the receiving side, copies them directly to the flash memory.

A risk with this technique is that because the firmware image is updated in real-time, the status of the firmware stored in memory can fall into an inconsistent state in case of failures (e.g. transmission errors). A countermeasure is to make all instructions that contain references to functions that are being updated, to halt till the update operation completes. Furthermore, when a new firmware update is released, multiple patches may be transmitted, each one replacing a specific part of the current firmware. Such a patch could also target the firmware code responsible for the updating; hence, code modifications must take place atomically, even when multiple patches are involved in the process.

As mentioned before, the code segments being updated must be suspended to avoid the node falling into an inconsistent state. However, code-halting till all patches are applied, can cause a significant delay, similar to that caused when a node reboot is required. A countermeasure is to minimise the required memory writes (called as \emph{atomic update set}) during the creation of an updated firmware. Two code patch strategies are proposed in~\cite{zhang_live_2016}: (i)~\emph{in-place patching} and (ii)~\emph{in-place patching with trampolines}.  

In~\emph{in-place patching}, all new and modified code parts are stored in the free space of the flash memory. The code insertion is performed by adding a jump instruction in the original code that has as target address, the location where new code has been inserted. Respectively, at the end of the inserted segment, another jump instruction exists that jumps back to the next instruction in the original image. Similarly, code deletion includes a jump from the address of the first deleted instruction to the instruction immediately following the last deleted one. Hence, the atomic update set contains all jumps added in the original code and not the modified parts that are transmitted, since the latter part of the code is placed at a safe place in the flash memory, where no conflicts can occur. This way, the size of the set is proportional to the number of the firmware parts required to be patched.

In~\emph{in-place patching with trampolines}, a code snippet, called as trampoline, is used for each patch. While the modified code is inserted in an unused memory location, like in the previous technique, the original code does not jump to these patches (at least not directly). However, a jump instruction in inserted in the original code to redirect the control to the corresponding trampoline. 
Initially, the trampolines return back to the instruction that follows the jump instruction in the original code. This is accomplished utilising a centralised approach that instructs all trampolines to jump back to the original code, so no conflicts can occur and hence the atomic set is empty. Afterwards, a \emph{base} variable, that all trampolines check to determine their return addresses, is modified and instructs the trampolines to jump to the corresponding patch (new code inserted in an unused memory location). Using this global variable as centralised approach and being able apply all patches in a single shot, results to minimal downtime, due to the small number of instructions in the update atomic set. Essentially, the atomic set consists of just the update of the  \emph{base} variable, the pointer that directs all trampolines to jump to the corresponding patch.

\subsubsection{Dynamic linking of modified firmware sections}
The use of a dynamic linker~\cite{dunkels_run-time_2006} at the node side was introduced as part of the first version of the Contiki OS~\cite{dunkels_contiki_2004}, an operating system for constrained IoT devices. In this OTAP scheme, only the modified section(s) need to be transmitted, since upon reception, the dynamic linker will re-link the image and load it again, replacing the previous one in the program flash. However, a limitation of this technique is that it requires an operating system or a sophisticated linker that can resolve the addresses properly. Moreover, the modified section is usually accompanied by the new symbol and relocation tables that will be used during the dynamic linking, increasing transmission overhead.

\subsubsection{Modules extraction} 
This technique was presented by the incremental firmware update algorithm \emph{MoRE}~\cite{park_non-invasive_2014}. In order to create the firmware image, many object files are linked together. The MAP file format is used, which is a text file format that presents the relative offset and length of each object in a binary file. The proposed method extracts binary fragments called modules from a firmware image. By comparing the modules extracted from two sequential incremental updates, it becomes feasible to detect which modules are modified. This was accomplished using modified versions of the \emph{R3diff}~\cite{dong_r3_2013} and \emph{RMTD}~\cite{hu_reprogramming_2009} image comparison algorithms for the delta calculations.  
In evaluation experiments, although \emph{MoRe} in most cases resulted to slightly larger data transmissions compared to the (modified) \emph{RMTD} and \emph{R3diff} algorithms, it performs better in terms of the required time to complete. It must also be noted that in \emph{MoRe}, there is no need to transmit extra information such as metadata and relocation-indirection tables, thus resulting to lower network overhead.

\subsubsection{Replaceable components}
The concept of \emph{replaceable components} was introduced in the \emph{Elon} reprogramming scheme~\cite{dong_elon_2010}, as a way to reduce the data code to be transmitted during a firmware update (for the TinyOS operating system). \emph{Elon} is based on the assumption that TinyOS kernel components (e.g. CTP, FTSP) are rarely updated, in contrast to the application components; hence, the programmer can define in the source code which components (code and data) can be updated in a future version. To do so, programmers have to use the appropriate annotations in the source code of the base firmware (e.g.~\emph{@replaceable}, \emph{@system}).

In \emph{Elon}, the replaceable components and system components are stored in different sections, \emph{.vdata/.vbss} and \emph{.vtext} for replaceable data and code, respectively. Initially, when the first (base) version of the firmware is created, the replaceable sections are stored in the flash memory and are not removed throughout node's lifetime, as they serve as a \emph{golden image} to be used for fail-safe operations. In subsequent firmware updates, the replaceable components that are received by a node, are stored in the RAM to avoid further flash memory access. In order for the linker to determine the base address of these sections, a two-phase linking process is required. During the first phase, the firmware is compiled and the size of these sections can be determined, since they do not involve external libraries that may be invoked after linking (external libraries cannot be replaceable). Given this, the linker is able to find the starting addresses of all sections in both the RAM and flash memory. Moreover, Elon utilises an indirection table that is called as \emph{jump table}. This is crucial, because the replaceable components can be relocated and the indirection table helps to mitigate the effect of these shifts.

Based on this technique, \emph{Elon} is able to update firmware without the need for a node reboot. A kind of software reboot process takes place, where, initially, the replaceable sections and the new jump table are stored in RAM, and then the control jumps to the initial address of the firmware image. This way, no kernel data initialisation is required, minimising the required downtime.

An advantage of \emph{Elon}, over other OTAP schemes, is that it does not require sophisticated OS support (e.g. dynamic linking). Moreover, it does not produce additional metadata e.g. indirection or relocation tables. However, a concern exists with regards to the size of RAM required to store modern and complicated firmware. Moreover, as the replaceable components are stored in RAM, they are not persistent and have to be re-transmitted in case of a hardware reset. Finally, in \emph{Elon} is assumed that the core system code and the libraries will not be updated in future firmware versions.

\subsection{Differencing algorithms}
\label{sec:diff_algo}

As described in the previous section, several techniques exist that preserve the similarity of the two firmware images, mitigating the effects of the function and variable shifts. When this process has been completed, it is desirable to minimise the required data that needs to be transmitted for the update of the IoT nodes. To make this feasible, the sending station (e.g a firmware server) that initiates the firmware update, uses a \textit{differencing} algorithm in order to find the common segments between two firmware images (the firmware that the nodes currently run and the updated one). Differencing algorithms can be of two types; either block-level or byte-level, depending on the granularity level they are able to detect matching segments.



The block-level algorithms (e.g.~\cite{jeong_node-level_2003, tridgell_efficient_1999}) split the firmware images into fixed-size blocks, aiming to detect non-common segments between the two images; hence, their accuracy is highly affected by the block size. On the other hand, the byte-level algorithms (e.g.~\cite{dong_r3_2013, hu_reprogramming_2009}) are able to find non-common segments between two firmware versions using blocks of variable lengths and can utilise more fine-grained approaches in order to achieve better accuracy, for example dynamic programming~\cite{hu_reprogramming_2009}. Regarding algorithms' performance, the block-level ones can detect a limited number of non-common segments, since they are not able detect those with size smaller than the size of a block. However, these algorithms typically have smaller time and memory footprint. Byte-level algorithms, on the other hand, can detect more non-common segments but typically require more time to complete.

With respect to the limitations of the IoT constrained nodes, it is evident that in order to utilise the available resources more efficiently, the required amount of transmitted data during a firmware update has to be minimised. This is achieved through a process referred as \emph{delta script generation}, which exploits two principles: (i) the nodes to be updated have already stored a previous firmware version in their flash memory, and (ii) the updates mostly introduce small modifications to the firmware binary code. The key idea of the delta scripts is to transmit only the parts of the firmware that have been altered, followed by instructions (for the node) regarding the local firmware reconstruction.


Hence, the common and non-common segments detected by the differencing algorithm, along with some special-purpose instructions, are encoded into commands of a \emph{delta script} that is transmitted to the receiving nodes. Based on this script and the current firmware code, each node is able to re-construct the new firmware version locally, executing the commands in the the script. It is evident that the delta script creator should be aware of: (i) the current firmware image a node currently executes, and (ii) and how the current image  is mapped in the internal memory, because the encoding highly depends on these two.

There has been an effort to create universal delta script formats, e.g. VCDIFF~\cite{korn_vcdiff_2002}, as well as some other custom ones~\cite{dong_r3_2013,dong_r2_2013} that support additional instructions in order to achieve a more efficient data encoding. Despite their differences, all proposed formats feature two common core operations with relatively standardised syntax: \emph{COPY} and \emph{ADD}. These two instructions, along with others reported in the literature, are presented below.

\begin{itemize}
   \item \textbf{\emph{COPY}}~\cite{panta_zephyr_2009}: This instruction is used to encode the segments of the new image that have been matched with others of the current version. The update module at the receiving side, upon interpreting the \emph{COPY} instruction, copies a segment with a pre-defined length from the current firmware image. Moreover, the starting address of the sequence must be provided. Once the update module has determined the length and the starting address of the copied segment, it copies and appends it to the new firmware image (that is reconstructed locally).
       
   \item \textbf{\emph{ADD}}~\cite{panta_zephyr_2009}: This instruction is used to encode the segments of the new image that do not match with other segments of the current version, and hence, they have to be fully transmitted. When the \emph{ADD} instruction is interpreted by the update module at the receiving side, the associated received sequence will be used (appended) to reconstruct the firmware image.
   
   \item \textbf{PAD}~\cite{ondrej_effective_nodate}: Pads the memory with a specific data unit.
   
   \item \textbf{RUN}~\cite{korn_vcdiff_2002}: Used for a more efficient transmission of repeated data. The update module copies the data that are associated with the instruction a finite number of times.
    
   \item \textbf{REPEAT}~\cite{panta_zephyr_2009}:  Used when specific patterns are detected in the data that is transmitted with small differences between them, that are introduced in a predictable and standardized manner.
\end{itemize}

Some of the most popular differencing algorithms are shown in Table~\ref{differencing_algorithms} and analysed in the following sections.

\begin{table}[ht]
\caption{Summary of differencing algorithms commonly used for OTAP schemes $n$: the combined length of the two firmware images in bytes}
\centering
\renewcommand{\arraystretch}{1.5}
\begin{tabular}{| L{0.12\textwidth} | L{0.10\textwidth} | L{0.08\textwidth} | L{0.08\textwidth} | }
\hline
\textbf{Algorithm} & \textbf{Type} & \textbf{Time complexity} & \textbf{Space complexity} \\
\hhline{|=|=|=|=|}
\hline
\textbf{FBC}~\cite{jeong_node-level_2003} & block-level & $O(n)$ & $O(n)$ \\ 
\hline
\textbf{Rsync}~\cite{tridgell_efficient_1999} & block-level & $O(n^2)$ & $O(n)$ \\ 
\hline
\textbf{RMTD}~\cite{hu_reprogramming_2009} & byte-level & $O(n^3)$ & $O(n^2)$ \\ 
\hline
\textbf{Hirschberg's trick}~\cite{mazumder_efficient_2013} & byte-level  & $O(n^2)$ & $O(n)$ \\ 
\hline
\textbf{R3diff}~\cite{dong_r3_2013} & byte-level & $O(n^3)$ & $O(n)$ \\ 
\hline
\textbf{DASA}~\cite{mo_efficient_2012} & byte-level  & $O(nlogn)$ & $O(n)$ \\ 
\hline
\textbf{DG}~\cite{kachman_optimized_2016} & byte-level &  $O(n^2)$ & $O(n)$ \\ 
\hline
\end{tabular}
\label{differencing_algorithms}
\end{table}

\subsubsection{Fixed block comparison (FBC)}
\emph{FBC}~\cite{jeong_node-level_2003} is the simplest method for comparing two firmware images, aiming to minimise the required data transmission for an update. This algorithm splits the two images into blocks and then compares each corresponding block. For each matching block pair, a \emph{COPY} instruction is inserted into the produced delta script, while the non-matching ones are transmitted along with the delta script. In order to encode the latter blocks, an \emph{ADD} instruction needs to be inserted in the delta script.

The main benefit of this technique is the low time and space overhead, as well as the ease of implementation. Moreover, it works well for small firmware changes, since only the altered blocks are transmitted. Nevertheless, it operates at block level granularity and is not able to detect a high number of common pairs, especially when the update includes excessive modifications.

\subsubsection{Rsync}
\emph{Rsync}~\cite{tridgell_efficient_1999} is an algorithm used by many incremental reprogramming schemes, (e.g. in~\cite{jaein_jeong_incremental_2004, panta_zephyr_2009}), in order to compute the common segments of two firmware images, initially developed for binary files exchange over low-bandwidth channels. This is a block-level differencing algorithm that splits the firmware images into fixed-size blocks, and then uses a sliding window with a size equal to the block size, to scan the two firmware images for detecting matching segments. Initially, a \{Checksum, MD4\} pair is calculated for each block of the current firmware image and then, the window traverses the new image, the \{Checksum, MD4\} pair of each window is calculated and lookups with the pairs of the current image are performed to detect potential matches. Like typical sliding window protocols, when a match is found, the window moves forward one block, otherwise it moves one byte, signing this block as unmatched. All  unmatched blocks are accumulated for transmission either when a next block is matched, or the current window reaches the end of the new image. 

Although Rsync is able to find subsequencies with a higher accuracy compared to \emph{FBC}, it still faces similar drawbacks, since its granularity depends on the window size used; thus, being not able to detect common segments with a size smaller than that of the window used.

\subsubsection{Reprogramming with minimal transferred data (RMTD)}
\emph{RMTD}~\cite{hu_reprogramming_2009} is a byte-level algorithm that aims to find the optimum combination of common sequences between two images, in order to minimise the number of transmitted bytes. Similarly to other differencing algorithms, RMTD finds the common segments between the two firmware versions. Nevertheless, a novelty of this algorithm, is that it utilises the partially reconstructed firmware image to detect matching segments. As the instructions in the delta script are executed at the receiving side sequentially, the new firmware is gradually rebuilt. Hence, for each segment of the new firmware, RMTD checks if it can also find matching segments in this partially rebuilt image.

RMTD uses a 2D matrix to record the pairs of the common bytes found for the two firmware images, with the comparisons performed in both forward and backward order to achieve higher accuracy. The result of this operation consists of two lists that contain the matching segments of the two images, as well as the matching segments between the partially reconstructed new image and the rest of the (new) image, respectively.

Once these two lists are computed, the algorithm finds the optimal combination of the \emph{COPY} and DOWNLOAD instructions to encode the common segments. Regarding the segments that can be encoded using \emph{COPY} instructions (found in any of the two lists described above), many such common sequences may correspond to the same memory addresses. When the respected \emph{COPY} instruction found in the delta script are executed, they will result to multiple writes of the same data to the same memory addresses; hence, these redundant writes lead to energy waste and also affect flash memory life span. To mitigate this issue, RMTD finds the optimal combination of \emph{COPY} instructions on the detected common sequences, using a dynamic programming approach. 



Finally, it must be noted that the algorithm's complexity depends on the size of the images, which makes it unsuitable for increasingly complex programs, as the time required to complete is substantially high. Moreover, in experiments conducted by other researchers (e.g.~\cite{mo_efficient_2012}), it was shown that RMTD crashes when the code size becomes too large (\~42 Kb), due to lack of memory.

\subsubsection{Hirschberg's trick}
Hirschberg's trick~\cite{hirschberg_linear_1975} is a method for computing the longest common sequences between two strings, while saving space, utilising a dynamic programming approach. A longest common subsequence ($LCS$) of two strings $X$=$x_{1}x_{2}x_{3}...x_{m}$ and $Y$=$y_{1}y_{2}y_{3}...y_{n}$ is a subsequence of both X and Y, whose length is the maximum possible. Let all prefixes of the two strings $X$ and $Y$ be \{$X_{1}, X_{2}, X_{3}, ...X_{m}$\} and \{$Y_{1}, Y_{2}, Y_{3}, ...Y_{n}$\}, respectively, where $X_{i}$ and $Y_{i}$ represent the prefixes that contain the first $i$ bytes of the corresponding string. Moreover, between two prefixes $X_{i}$ and $Y_{j}$, there may be multiple longest common prefixes but all of them will have equal length; hence, if we denote the length of the $LCS$ of these prefixes as $C(i,j)$, its dynamic formulation is as follows:

\begin{equation*}
C(i,j)=\begin{cases}
          0  \quad &\text{if} \, i=0 \ or \ j=0\\
          C(i-1,j-1)+1 \quad &\text{if} \, x_{i}=y_{j}  \\
          max(C(i-1,j),C(i,j-1)) \quad &\text{if} \, x_{i} \ne y_{j} \\
     \end{cases}
  \end{equation*}
  
Using the above formulation, the common subsequences between the prefixes of two images can be found in order to compute the delta script. Based on Hirschberg's trick, the authors in~\cite{mazumder_efficient_2013} proposed a differencing algorithm to detect common firmware subsequences. Hirschberg also presented a modified version of this algorithm, which follows a divide-and-conquer approach and is able to compute the \emph{LCS} of two strings in O(min(m,n)).

\subsubsection{R3diff}
\label{sec:r3diff}
\emph{R3diff}~\cite{dong_r3_2013} is a byte-level comparison algorithm that complies with the overall design of the R3 OTAP scheme. Initially, the algorithm computes the hash values for every three continuous bytes of the current image. Three bytes were chosen as the lowest level of granularity because copying smaller byte segments (e.g. 2-byte long), is not more beneficial than adding them, due to the overhead of the \emph{COPY} instructions that comes with additional parameters in the delta script. 

In order to compute the optimal delta for transmission, the authors use the $opt_{i}$ annotation that represents the minimum delta script size that needs to be transmitted, in order the first $i$ bytes of the update image to be reconstructed by the receiving node. Moreover, a \emph{findK} method is used for each such prefix (first $i$ bytes) that returns the smallest index $k$, so that the subsequence [$k$,$i-1$] is a common segment (found both in the new and current images). Furthermore, two additional notations are used by the algorithm, $opt_{i}^A$ and $opt_{i}^C$ that represent the minimum number of bytes that need to be transmitted in the delta script in order to reconstruct $i$ bytes, having as last instruction an \emph{ADD} or a \emph{COPY} one, respectively. Hence, in order to reconstruct the first $i$ bytes, $opt_{i}=min(opt_{i}^A, opt_{i}^C)$, omitting the overhead of \emph{COPY} instructions, as encoded in the formulations below.

To compute the optimal $opt_{i}$, the algorithm follows a recursive approach, starting with $opt_{0} = 0$, iterating over all possible prefixes, till the final one that corresponds to the new image. For each prefix, the algorithm computes $opt_{i}^A$ and $opt_{i}^C$, and also runs the \emph{findK} method to check if the prefix can be encoded in a \emph{COPY} instruction using a common segment. If no such common segment is found, $opt_{i}^C$ is set to a large integer, so that is not selected over $opt_{i}^A$. The formulations of $opt_{i}^A$ and $opt_{i}^C$ are as follows:

  \begin{equation*}
    opt_{i}^A=min(opt_{i-1}^A+1,opt_{i-1}^C+\alpha+1)
  \end{equation*}

  \begin{equation*}
    opt_{i}^C=\begin{cases}
          LARGE\_INTEGER  \quad &\text{if} \, k>i-1\\
         opt_{k}+\beta \quad & \, otherwise  \\
     \end{cases}
  \end{equation*}
, where $\alpha$ is the additional overhead, imposed by each \emph{COPY} instruction and $\beta$ the overhead of an \emph{ADD} instruction.

\subsubsection{An efficient differencing algorithm based on suffix array (DASA)}
\emph{DASA}~\cite{mo_efficient_2012} is a differencing algorithm that focuses on minimising the space and time complexity for computing the optimal delta script. In order to accomplish this, it utilises an efficient data structure, called \emph{suffix array} (\emph{SA})~\cite{dementiev_better_2008}. \emph{SAs} can be used for data processing, as well as for data compression operations. Initially, the algorithm combines the two firmware images in a \textbf{\$-\#}-padded extension format, using them in reverse order. For example, two strings $S_{1}$ = "cdn" and $S_2$ = "ngtv" will result to $T$ = "ndc\#vtgn\$". 

Once the extension is created, the doubling algorithm~\cite{dementiev_better_2008} is used to compute the \emph{SA} by sorting all possible suffixes in ascending order and for each suffix, stores its starting index in $T$. After the computation of \emph{SA} is completed, it is used by \emph{DASA} as input to create the height array in time complexity of $O(nlogn)$. The height array contains the length of the longest common prefix of each suffix with the next one in sorted order. Having computed the height array, one is able to get the LCP (longest common prefix) of any two suffixes in linear time.

Next, the algorithm computes the optimal delta script, annotating as $opt_{i}$ the optimal delta script size to reconstruct the first $i$ bytes of the new image ($opt_{0}=0$). For each prefix of the new image, \emph{DASA} uses the \emph{findK} method (discussed in Section~\ref{sec:r3diff}) to find the largest common segment that can be used to encode the prefix using \emph{COPY} instructions. Moreover, \emph{DASA} utilises the formulation of $opt_{i}^A$ and $opt_{i}^C$, similarly to \emph{R3diff}, in order to find the $opt_{i}$. Therefore, \emph{DASA} iterates over all possible prefixes to find the optimal combination of \emph{ADD} and \emph{COPY} instructions.

The experimental evaluation shows that \emph{DASA} outperforms \emph{Rsync} in terms of the delta script size, which is an expected, as \emph{Rsync} is a block-level algorithm. Despite several improvements introduced by the \emph{Rsync} developers, \emph{RMTD} and \emph{DASA} still have superior performance. Furthermore, \emph{DASA} has better performance than \emph{RMTD} both in time and space domains and this stands true especially when the new image is relatively large in size.

\subsubsection{Delta generator (DG)}
Many differencing algorithms assume abundance of spare memory~\cite{hu_reprogramming_2009} at the node side for firmware reconstruction, something not always true due to the constrained nature of several IoT node types. In~\cite{kachman_optimized_2016}, the authors propose a differencing algorithm, known as \emph{DG}, for nodes that lack external memory. The algorithm by exploiting the fact that two sequential firmware versions usually share many common parts, places the two images side-by-side and executes an XOR operation between the corresponding bytes, aiming to reveal the sequences of the non-matching bytes. The matching sequences can easily be encoded using \emph{COPY} instructions in the delta script and the non-matching ones can be further broken down into matching and non-matching subsequences, to achieve delta script size minimisation. To make this feasible, each non-matching sequence is checked against the current firmware to find matching subsequences.

The common subsequences found are reconstructed using \emph{COPY} instructions. This algorithm has $(O(n+m)$ space complexity and $(O(nm)$ time complexity, where \emph{n} is the size of the current firmware image and \emph{m} the size of the non-matching segments. A comparison of \emph{R3diff} and \emph{DG} was conducted in~\cite{lehniger_impact_2019}, using various image sizes and code shifts. The authors inferred that \emph{DG} outputs significantly smaller delta scripts than \emph{D3diff} for small-sized images but this does not stand true, as more data and code is shifted. Moreover, the authors found that the number of \emph{ADD} instructions in the delta script gets smaller, as code shifts increases. The authors conclude that \emph{DG} is not able to provide optimisation for a high number of small changes. Instead, it generates a number of \emph{ADD} instructions that encode regions with a few bytes for each non-matching segment. When the code shifts increase, these non-matching segments expand together and are merged under one common \emph{ADD} instruction. This results to a larger delta script with fewer \emph{ADD} instructions.

\begin{table*}[ht]
\caption{Summary of OTAP schemes}
\centering
\renewcommand{\arraystretch}{1.5}
\begin{tabular}{| L{0.13\textwidth} | L{0.13\textwidth} |  L{0.1\textwidth} | L{0.15\textwidth} | L{0.12\textwidth} | L{0.09\textwidth} | L{0.11\textwidth} | }
\hline

    \textbf{OTAP scheme} & \textbf{OS/platform} & \textbf{Update type} & \textbf{Firmware similarity} & \textbf{Differencing algorithm} & \textbf{Live update} & \textbf{Dissemination protocol} \\
\hhline{|=|=|=|=|=|=|=|}
\textbf{Elon}~\cite{dong_elon_2010} &
TinyOS, TelosB &
Incremental (replaceable components) &
Replaceable components &
- & 
Yes  &
Deluge \\
\hline

\textbf{R2}~\cite{dong_r2_2013} &
TinyOS, TelosB &
Incremental (delta script) &
R2sim (Relocatable code) & 
RMTD &
No  &
Stream \\
\hline

\textbf{R3}~\cite{dong_r3_2013} &
TinyOS, TelosB &
Incremental (delta script) &
R3sim (Relocatable code) & 
R3diff (Based on DASA) &
No  &
Stream \\
\hline

\textbf{MoRE}~\cite{park_non-invasive_2014} &
NanoQplus OS, Mango-Etoi board &
Incremental (Modified modules) &
- & 
RMTD, R3diff &
No  &
- \\
\hline

\textbf{Zephyr}~\cite{panta_zephyr_2009} &
TinyOS, Mica2 &
Incremental (delta script) &
Indirection tables & 
Rsync &
No &
Stream \\
\hline

\textbf{Hermes}~\cite{panta_hermes_2009} &
TinyOS, Mica2 &
Incremental (delta script) &
Indirection tables, global variables pinning & 
Rsync &
No  &
Stream \\
\hline

\textbf{In-place patching}~\cite{zhang_live_2016} &
TinyOS, MSP430FR5739 &
Incremental (Patches) &
In-place patching (with or without trampolines) & 
Rsync or Zephyr &
Yes &
- \\
\hline

\textbf{Qdiff}~\cite{shafi_no-reboot_2012} &
TinyOS, IRIS mote &
Incremental (delta script) &
Slop regions, RAM layout modification & 
Google-guava API, Google diff-match-patch &
No &
- \\
\hline

\end{tabular}
\label{tbl:OTAP_deltas}

\end{table*}

\subsection{Dissemination protocols}
Traditional data dissemination protocols (e.g.~\cite{stann_rmst_2003, kulkarni_infuse_2006, naik_sprinkler_2005}), initially designed for WSNs, are not suitable for firmware dissemination in IoT networks for a number of reasons. First, the size of the update image is typically larger (in the order of kilobytes) than that of the commonly transmitted data, while the corresponding protocols have been specifically designed to propagate small-size packets with a low packet rate. Furthermore, during the update process, the network nodes usually store the update image and then act as sources, which is not a typical behaviour in data dissemination protocols. Finally, while the flow of data transmissions in a WSN is bidirectional, including operation information from sensors and commands toward actuators, etc., the flow of the update image is one-way, from the base station to the network nodes. 

Hence, protocols designed for disseminating firmware updates in WSNs focus on the efficient distribution of the new firmware code from a central firmware repository server to the nodes. These protocols have to cope with the unreliable nature of the wireless medium nodes communicate, employing suitable mechanisms for the provision of a reliable firmware update process. Regarding this, the nodes should be able to provide a form of feedback, indicating the correct reception of update (network) packets, as well as to request the retransmission of lost ones.

The simplest method to disseminate a new firmware image is by flooding, where the firmware server broadcasts the new firmware code to its connected IoT nodes, and the latter further broadcast it to their neighbors in an epidemic fashion. However, redundant transmissions during flooding should be minimised, as they can easily deplete nodes' battery and can cause the \emph{broadcast storm problem}~\cite{tseng_broadcast_2002}, where overlapping radio signals result to increased contention and packet collisions.

In order to minimise the number of messages required during a firmware update, several protocols are proposed that take into consideration the underlying network topology. For example in~\cite{alagar_reliable_1995}, a broadcast protocol is presented that provides reliable data propagation within the network, by splitting the nodes into several clusters. As wireless networks suffer from problems like collisions and the hidden terminal problem, in order to minimise the required transmissions, most dissemination protocols instruct nodes to aggregate data prior to transmission to their neighbors. Moreover, they typically use a three-way handshake pattern to establish a communication channel between the firmware repository server and the nodes. Initially, the available sources advertise the available firmware version by broadcasting a message to their neighbors. Since some nodes may receive multiple advertisements of this type, they select a specific node as the firmware source based on some heuristics, and then broadcast a request packet directed towards this selected source. Once the request packet is received (by the source), the actual data transmission begins. When a node has partially or fully received an update image, it can in return broadcast an advertisement, indicating that it can now offer the firmware update as a source. The dissemination protocol can also be based on a subscription approach (e.g.~\cite{stathopoulos_remote_2004}), where the nodes of the network subscribe to sources in order to receive firmware updates. This in return, results to additional overhead for the sources, which are also nodes of the network, as they have to track all subscriptions.

Finally, most update dissemination protocols require a form of feedback from the nodes to the sources in order to validate packet correct reception. This can be accomplished either by ACKs or selective NACKs (negative ACKs). In the first approach, the node upon successfully receiving a packet, transmits a short-length ACK to the sender. If the source has not received an ACK from a node within a predefined time interval, it will re-transmit the packet. Typically, the source will try to transmit lost packets several times and will withdraw after a number of failed attempts. As every packet has to be ACKed separately, network implosion problem is possible in case of a large number of missing packets. To deal with this problem, several contributions use the NACK option, where a NACK is sent from a node to the source, only if a packet has not been successfully received, effectively reducing the number of control packets that need to be broadcasted during the firmware update process.


Several update dissemination protocols are proposed with the characteristics stated above. Some of them also use \emph{pipelining} (e.g. \cite{chlipala_deluge_2004,hagedorn_rateless_2008,dong_efficient_2014}), a method that allows the parallel transfer of data within the network, thus achieving better performance in terms of the time required to transmit a whole firmware image. Initially, the image is split into segments and dissemination is performed segment-by-segment. When a node completely receives a segment, it can become a source and can further disseminate it to its neighbors. 

In order to validate or compare the performance of the dissemination protocols, authors typically utilise simulation frameworks (e.g. TOSSIM~\cite{levis_tossim_2003}, EmStar~\cite{elson_emstar_2003}) or even use real world sensor deployments as empirical testbeds (e.g.~\cite{levis_trickle_2004, panta_stream_2007, kulkarni_mnp_2005, huang_cord_2008}). The latter can provide more accurate results, as the experiments are usually conducted in various realistic indoor and outdoor environments. However, in order to stimulate congestion and collisions, the authors are required to use a large number of networking devices, drastically increasing validation cost. To this end, the authors use simulation frameworks, where they can simulate network topologies along with links' quality, thus being able to increase the number of the nodes and observe the scalability and behaviour of the proposed protocol.

Some of the most commonly used protocols in OTAP schemes are presented in the following sections (summarised in Table~\ref{tbl:dissemination_summary}).


\begin{table*}[ht]
\caption{Dissemination protocols for Over The Air Programming}
\centering
\renewcommand{\arraystretch}{1.5}
\begin{tabular}{| L{0.11\textwidth} | L{0.11\textwidth} |  L{0.08\textwidth} | L{0.11\textwidth} | L{0.14\textwidth} | L{0.15\textwidth} | L{0.15\textwidth} | }
\hline
    \textbf{Protocol} & \textbf{OS/platform} & \textbf{Pipelining} & \textbf{Dissemination hops} & \textbf{Encoding} & \textbf{Feedback /Reliability strategy} & \textbf{Experimental validation} \\
\hhline{|=|=|=|=|=|=|=|}

\textbf{Trickle}~\cite{levis_trickle_2004} &
TinyOS, Mica-2 &
No &
Multi-hop &
Full image & 
NACK-based  &
Simulated (TOSSIM) \\
\hline

\textbf{Deluge}~\cite{chlipala_deluge_2004} &
TinyOS, Mica-2 &
Yes (pages) &
Multi-hop & 
Full image (can also support incremental updates) &
NACK-based  &
Simulated (TOSSIM) \\
\hline

\textbf{Rateless \& ACKLess Deluge}~\cite{hagedorn_rateless_2008} &
TinyOS, Tmote Sky &
Yes (pages) &
Multi-hop & 
Full image (can also support incremental updates) &
FEC  &
Simulated (TOSSIM) and test-bed \\
\hline

\textbf{MOAP}~\cite{stathopoulos_remote_2004} &
TinyOS, Mica-2 &
No &
Multi-hop & 
Full image (can also support incremental updates) &
NACK-based  &
Simulated (EmStar) and test-bed \\
\hline

\textbf{MNP}~\cite{kulkarni_mnp_2005} &
TinyOS, Mica-2 and XSM &
Pipelined (Segments) and non-pipelined  &
Multi-hop & 
Full image (can also support incremental updates) &
NACK-based  &
Simulated (TOSSIM) and test-bed \\
\hline

\textbf{XNP}~\cite{crossbow_technology_inc_2003_mote_nodate} &
TinyOS, Mica-2 &
No &
Single-hop &
Full image & 
Queries originated by the base station  &
Test-bed (Mica-2) \\
\hline

\textbf{CORD}~\cite{huang_cord_2008} &
TinyOS &
No &
Core-based &
Full image & 
No feedback  &
Test-bed \\
\hline

\textbf{ACDP}~\cite{dong_efficient_2014} &
TinyOS, TelosB &
Yes (pages) &
Multi-hop & 
Full image &
NACK-based (Unicasted)  &
Test-bed \\
\hline

\textbf{Stream}~\cite{panta_stream_2007} &
TinyOS, Mica2 &
Yes (pages) &
Multi-hop & 
Full image (can also support incremental updates) &
ACK-based &
Simulated (TOSSIM) and test-bed \\
\hline

\end{tabular}
\label{tbl:dissemination_summary}

\end{table*}

\subsubsection{Trickle}
\emph{Trickle}~\cite{levis_trickle_2004} is an update dissemination algorithm built for the TinyOS  and the Mica-2 motes~\cite{hill_mica_2002}, following a "polite gossip" approach to propagate the new firmware code throughout the nodes of a network. In \emph{Trickle}, each node periodically broadcasts an announcement that contains the current firmware version this node executes, informing others about a potential update. Firmware update propagation takes place if a node overhears that another node can provide a newer firmware version, or if a node receives a broadcast by another one indicating  that it executes an older version. This "polite gossip" approach makes \emph{Trickle} robust and scalable, able to operate in various networking environments (sparse, dense network topologies). Moreover, when a node overhears that a neighbor broadcasts metadata for an outdated firmware version, it broadcasts the code of the newer version, initiating the update of the outdated node.

Each node breaks time into fixed-size intervals and at a random point within an interval, it broadcasts announcements (metadata) that inform other nodes about the firmware version it can provide. However, if a node has overheard several other nodes broadcasting the same metadata, it stays quiet, since acting as a source can cause redundant transmissions, wasting network resources. When a node overhears such an announcement, there are two possible cases: (i) the node executes a newer firmware version than the one broadcasted, and (ii) a node executes an older version than the one broadcasted.

In the first case, the node will respond by broadcasting its (newer) code, while in the second case, the node will trigger a firmware update process by broadcasting its (old) firmware version, so the node that initiated the "gossip", as soon it receives this message, it will broadcast its (new) code. Moreover, instead of using a fixed transmission rate per node, \emph{Trickle} dynamically regulates it by considering network density in order to achieve the desirable communication rate, permitting specific number of transmissions in each interval and for each node.

Network nodes must, in advance, synchronise the time intervals for their broadcasts, otherwise \emph{Trickle} could suffer from the short-listen problem; some nodes of the network exchange metadata immediately after the starting of their (time) interval period, listening for a short period of time, before any other node is able to broadcast its metadata. This can result to redundant transmissions because a node may never hear some other broadcasts and hence, will not suppress its transmissions. \emph{Trickle} overcomes this challenge by requiring nodes to fall into a listen-only state under which, they cannot broadcast any metadata. The other half of the time interval is then available for each of the nodes to broadcast its metadata.

Finally, \emph{Trickle} was one of the first developed update dissemination protocols and since then, it has been adopted as the basis for many dissemination protocols (e.g. Deluge~\cite{chlipala_deluge_2004, kulkarni_mnp_2005}), with many contributions aiming to further optimise it (e.g.~\cite{djamaa_optimizing_2015, ghaleb_e-trickle_2015}).


\subsubsection{Deluge}
\emph{Deluge}~\cite{chlipala_deluge_2004} was built for the Mote-2 devices as the default network reprogramming protocol for TinyOS. \emph{Deluge} employs a negotiation mechanism based on \emph{Trickle}, however with an increased performance, as it provides pipelined firmware  dissemination; hence, when a node has received a chunk (page) of the update image, it can also act as a source for it, serving its neighboring nodes. 

A firmware image prior to its dissemination, is split into pages, and each such page is further split into fixed-size packets that fit to the maximum packet size of the TinyOS network stack. Using \emph{Trickle}, nodes periodically advertise the pages of a firmware version they can provide through broadcast packets, while other nodes send requests for pages they are missing and are willing to receive. The available pages a node holds for a specific firmware version, are represented by a bit vector that is carried by the advertisement packets. \emph{Deluge} enforces a sequential transmission of the pages, so for a node to request a missing page, it is required to have successfully received all previous ones.

When a node receives an advertisement packet and infers that there is a new firmware version available, it first finds the lowest numbered page that it needs to receive. In most cases, where the firmware image binary has been completely changed, this page will be the first one. Once this page is determined, the node waits for a predefined time interval to receive further advertisements transmitted by neighboring nodes and to decide which of them can provide the specific page. When this period is up, the node heuristically selects one of the available source nodes and  transmits a request packet that indicates the page and the packets within the page that it wants to receive.

The selection of the most appropriate firmware source node for a specific page is a challenging task, as the total energy consumption and network bandwidth utilisation should be minimised. Heuristics used by \emph{Deluge} aiming to address these challenges, are as follows:

\begin{itemize}
    \item A node requests data from the source node that transmitted the most recent advertisement packet.
    \item A node requests data from the nearest source node. 
    \item A node requests data from the node farthest from the source, in order to inflate the overall spatial multiplexing.
    \item A node requests data from the node nearest to the source node.
\end{itemize}
The latter heuristic was used only for comparison to others, since it has many disadvantages. For example, neither it tries to promote high quality links, nor tries to improve the spatial multiplexing for a better data propagation performance.


Regarding the feedback mechanism of \emph{Deluge}, it is not explicitly based on ACKs or NACKs, because such packets are not transmitted but the protocol itself is described as NACK-based, because, by default, nodes re-request packets either not received, or corrupted. Additionally, in order to decrease energy consumption and excessive bandwidth utilisation, \emph{Deluge} uses a self-organised suppression mechanism that allows each node to deactivate itself, based on packets it overhears; for example, it does not transmit a page request if, in the meantime, it overhears the same request sent by a different node.

A disadvantage of \emph{Deluge} is that it requires the nodes' radio to be always turned on, resulting to an increased energy consumption. Moreover, the protocol does not support fault detection or recovery mechanisms. The authors proposed an optimisation based on forward error correction (FEC), using the \emph{digital fountain approach}~\cite{byers_digital_1998}, a method for efficient transmission of bulk data by heterogeneous nodes.

\subsubsection{Rateless and ACKLess Deluge}
\label{rateless}
\emph{Rateless Deluge}~\cite{hagedorn_rateless_2008} is a firmware update dissemination protocol based on \emph{Deluge}, however introducing several modifications to its propagation mechanism, aiming to enable rateless transfer of the firmware image and to reduce the overhead due to lost packets' retransmissions. Using a \emph{rateless coding} approach, the authors manage to minimise the amount of the control messages required. Additionally, the nodes do not need to explicitly specify which packets need to be retransmitted, as they only need to specify the number of packets successfully received.

In order to construct rateless codes, the authors utilised the theory of random linear codes (RLCs)~\cite{ho_random_2006}. A file $X$ to be transmitted is split into $k$ distinct segments that are encoded into $m>k$ messages. Each encoded message is computed as the weighted sum of all $k$ segments, $Y_i=\sum_{j=1}^{k}\beta_{i,j}X_j $. Hence, in order to compute each  message $Y_i$, a weight vector $\beta_i$ (coefficient) of size $k$ must be chosen that contains the weight of each segment for that specific message. These values should be adjusted so that the  coefficients of different encoded messages are linearly independent with high probability, so a node would only need to receive $k$ $Y_i$s in order to solve the corresponding system of linear equations and determine the initial file $X$. 

\emph{Rateless Deluge} follows the same approach, splitting the firmware image into pages that are further divided into packets, and then, encoding each packet using \emph{RLC}. Respectively, a receiving node stores the encoded packets in its memory, and once the number of received packets has reached the page size, it proceeds to decoding using the \emph{Gaussian elimination process} (GEP). If decoding is successful, the linear equation system is solved and the specific page is stored in the flash memory of the node, while it is requesting the next page.

However, if the received packets are linearly dependent, the process of \emph{GEP} fails. In this case, the node discards the linearly dependent packets and waits for some time to receive more packets, and then starts \emph{GEP} again. In contrast to \emph{Deluge}, if some packets are lost, the receiving node informs the firmware source about the lost packets, without specifically identifying them, thus reducing the required feedback overhead. To further improve performance, \emph{Rateless Deluge} exploits the fact that the pages are requested in an incremental fashion and pre-codes the next page.

\emph{ACKLess Deluge} adopts the changes introduced by \emph{Rateless Deluge} (on top of \emph{Deluge}), but its main goal is to further reduce the need of retransmissions by employing an FEC algorithm, which operates at the packet level and appends extra encoded information to avoid additional control messages and retransmissions. The amount of redundant information is proportional to the calculated loss probability of each distinct receiving node.

\subsubsection{Multihop Over-the-Air Programming (MOAP)}
\emph{MOAP}~\cite{stathopoulos_remote_2004} is another multi-hop OTAP dissemination protocol, specifically designed for the Mica-2 motes, executing TinyOS. The goal of \emph{MOAP} is to minimise RAM usage and energy consumption during a firmware update. This protocol does not offer pipelining, since for a node to become a firmware source, it has to receive the whole image in advance. The firmware update image is built using the standard TinyOS tools and is then split into a number of segments, with each segment transmitted using a single network packet. \emph{MOAP} dissemination mechanism is called \emph{Ripple} and, unlike other dissemination algorithms, it avoids network flooding by selective imagine forwarding to other nodes, also utilising a sliding window protocol for the identification of lost packets. 

\emph{MOAP} uses a publish-subscribe mechanism for image dissemination in a neighbor-by-neighbor node fashion. A node that has received the full firmware image and can become a firmware source, advertises information regarding the type and version of the firmware image it already holds, through a \emph{PUBLISH} message that is broadcasted periodically. Interested nodes transmit \emph{SUBSCRIBE} messages in order to subscribe for a firmware update. As soon as a source node receives such as message, it waits for some amount of time to receive any further subscriptions (in order to suppress duplicate transmissions) and then provides the new firmware image to the subscribed nodes, which can further become source nodes. These nodes wait again for some amount of time to receive any subscriptions, and if not, they reboot and the firmware update takes place. Eventually, all nodes are programmed with the new firmware version.

\subsubsection{Multi-hop network reprogramming (MNP)}
\emph{MNP}~\cite{kulkarni_mnp_2005} is a multi-hop reprogramming protocol, designed for the Mica-2 and XSM~\cite{prabal_dutta_design_2005} nodes running TinyOS. This protocol aims to reduce network collisions occurring during firmware dissemination by proposing a source node selection algorithm, which guarantees that at any given time, and for each neighborhood, at most one node can act as a firmware source. Additionally, \emph{MNP} improves propagation performance by supporting pipelined dissemination, splitting the firmware image into fixed-size chunks and permitting nodes to act as sources for the pages they have already received. 


Each source node has a unique identifier (ID) and maintains a \emph{ReqCtr} value that indicates the number of its receiving nodes (the nodes that have transmitted a request packet to this specific source node at least once). This value is incremented by one each time the source node receives a new request. At random intervals, each node broadcasts an announcement that contains its ID and \emph{ReqCtr}, as well as the firmware version ID. 
When a neighbor receives this announcement, it checks if interested in this new code and broadcasts a request, which contains the \emph{ReqCtr} and the ID of the corresponding source node. Using this technique, nodes unable to receive the initial source node announcement due to network problems (e.g. hidden terminal), can become aware of other nearby source nodes.  

Receiving nodes upon the reception of broadcast announcements, send replies back to the corresponding source nodes that increase (by one for every reply) their \emph{ReqCtr} values. The node with the highest \emph{ReqCtr} value is selected as a source node that can later disseminate the firmware to the interested receiving nodes. The source node, after the completion of a firmware update, enters into a sleep state for some amount of time, to enable a uniform load distribution, as other nodes can now have the chance to become source nodes. As soon as a new source node is selected, it broadcasts a \emph{StartDownload} message to inform potentially interested receiving nodes that has available firmware image to provide.     

\subsubsection{XNP}
\emph{XNP}~\cite{crossbow_technology_inc_2003_mote_nodate} is the earliest network reprogramming protocol hosted within the TinyOS operating system for the support of the Mica-2 motes. \emph{XNP} transmits the whole firmware image (no support for incremental programming) and is able to disseminate it from a firmware server to only its single-hop nodes. 

Initially, the server splits the firmware binary code into multiple packets that are then broadcasted one-by-one. Single-hop nodes able to receive these packets, store the carried information in their external memory. As packets can be lost, nodes request any lost packets until the entire binary code is correctly received. To accomplish this, \emph{XNP} at the firmware source node checks the successful delivery of each packet by querying the receiving node. When the image is completely received, \emph{XNP} uses the bootloader to copy the code to the flash memory and restarts the node~\cite{wang_two_2005}.

A major drawback of \emph{XNP} is that it does not support OTAP in multi-hop networks. Moreover, each time an update is required, the whole firmware image needs to be transmitted, as there is no delta mechanism supported. \emph{XNP} also occupies a significant portion of the program memory in the (constrained) nodes. Finally, during the download of the update image, the application is halted and the \emph{XNP} module that is wired into the firmware image executes. This introduces extra overhead in terms of time, proportional to the network latency.

\subsubsection{COre based Reliable Dissemination (CORD)}
\emph{CORD}~\cite{huang_cord_2008} is a reliable bulk data dissemination protocol that mainly targets energy consumption minimisation during the firmware update process. To achieve this, it follows a different approach compared to the other dissemination protocols presented in this paper, which due to their epidemic nature, propagate the update in a neighborhood by neighborhood fashion, employing also a three-way handshake pattern (advertise-request-data), resulting to a vast amount of control data. On the contrary, \emph{CORD} follows a core-based two-phase approach, similar to the one used in Sprinkler~\cite{naik_sprinkler_2005} and Garuda~\cite{park_scalable_2004}.

The authors claim that it is possible to identify \emph{reliable links} that have a constant low packet loss rate. Transmitting data over these links, can result to less corrupted/lost packets and subsequently less control messages. During the first phase of \emph{CORD}, a subset of nodes that are interconnected though reliable links is identified, forming an \emph{approximate minimum dominating set}~\cite{guha_approximation_1998}. These nodes are selected as the \emph{core nodes} using Cheng's single leader algorithm~\cite{cheng_virtual_2006}. Subsequently, each network node can either be a \emph{core node} or an immediate neighbor of such a node. Once the \emph{core nodes} are selected, the update image is propagated from a firmware source to the \emph{core nodes} through reliable multi-hop forwarding. The firmware image is initially split into pages and pipelining is used for increased performance. Once all \emph{core nodes} receive the update image, \emph{CORD}'s second phase begins where the image is disseminated to the rest of the nodes. Furthermore, \emph{CORD} uses a sleep scheduling schema to instruct nodes in the network that do not receive or transmit any data, to turn off their radios in order to reduce energy consumption.


\subsubsection{Adaptive Code Dissemination Protocol (ACDP)}

\emph{ACDP}~\cite{dong_efficient_2014} has been developed for TelosB nodes, executing TinyOS and aim to minimise the number of packets that need to be transmitted during the update image dissemination, providing reliability, low energy consumption, balanced traffic in the network and rapid propagation. Similarly to \emph{Rateless Deluge}, \emph{ACDP} employs random linear codes for efficient and robust data transmission.

Prior to dissemination, the update image is split into different pages and each page is further divided into a fixed number of packets. This new code image will be distributed by a single node, whereas the intermediate nodes will forward the pages they have received to their neighbors, incrementally, to ensure that the update will finally reach all network nodes. 
Additionally, at any given time, only one page is loaded into the RAM of a source node for transmission. When the node finishes with the transmission of a single page, the allocated RAM is released and it enters a sleep state, so that other nodes can transmit previous pages.

In order to encode the packets of a  page, \emph{ACDP} is based on \emph{RLC}, producing a linear combination of \emph{M} packets of that given page. The receiving node uses the Gaussian elimination method to find the original page, which is a relatively efficient task, as described in \ref{rateless}. Hence, when a node has received enough packets, it attempts running the Gaussian elimination method, aiming to decode the initial packets of the corresponding page. If decoding is successful, the receiver chooses a sliding window of size \emph{N}, depending on the number of its neighbors, and will encode \emph{N} packets, computing their linear combination. This way, the node will send out \emph{M/N} encoded packets into the network, achieving better load balancing within the network. The reason the size of the sliding window is proportional to the number of the neighbors is because when a node has many neighbors it can receive enough encoded packets to decode, without losing reliability. Finally, nodes broadcast NACKs  when a request has not been served within a pre-defined amount of time. 

\subsubsection{Stream}
\emph{Stream}~\cite{panta_stream_2007} is an update dissemination protocol that aims to minimise the data transmission overhead by pre-installing the module responsible for the firmware update in nodes' flash memory as a distinct image. Hence, at any given time, each node stores two images: (i) firmware image, (ii) reprogramming protocol image.

The application image contains the actual firmware image and a limited reprogramming code (Stream-AS), whose purpose will be discussed below, whereas the reprogramming protocol image contains the code for the reprogramming/dissemination protocol (Stream-RS) and its purpose is to enable the firmware code update that is pre-installed in all network nodes.

When a new update is released, all nodes are instructed to reboot to the stream-RS, so to permit the dissemination of the new firmware. To achieve this, a command is injected in the network from the base station that instructs the nodes to switch to the Stream-RS image. Once a node receives this command, it broadcasts this further to its neighbors and then reboots, using the functionality provided by Stream-AS, which is included in the image that it is currently running. When all nodes receive the reboot command, they run the Stream-RS image and the dissemination of the new update, augmented by the corresponding Stream-AS, is permitted.

Furthermore, Stream-RS follows a three-way-handshake approach for the actual dissemination of the firmware that is based on the \emph{Deluge} protocol. Each node contains a list of its neighbors that have transmitted firmware requests. When all these neighbors have successfully received all pages requested, the node is able to reboot from the application image, which now contains the updated firmware. After a certain period has elapsed, all nodes switch back to the application image, thus all get updated with the new image.   

\section{over-the-air programming security}
Although IoT networks are often used in critical infrastructures, they are also known for their weak security if not deployed properly with security in mind. Similarly, the firmware update mechanism itself can also be a major attack vector, if not designed appropriately. Due to the broadcast nature of the wireless medium, an adversary can launch both \emph{external} and \emph{insider} attacks. In external attacks, the attacker does not control any network node; however, he can inject forged packets, launch replay attacks or even impersonate nodes. He can also launch DOS attacks injecting a vast amount of messages and exploiting the weakness of the dissemination protocol.

In insider attacks, the adversary has managed to compromise a node and instructs it to intercept sensitive information, drop packets, inject false data, exploit weaknesses of the protocol, etc. Due to the  epidemic nature of the update dissemination protocols, an adversary can gain complete control over the network, by compromising a single node and then leveraging the reprogramming mechanism to distribute malicious code in every reachable node.

For these reasons, the authenticity and integrity of the firmware image have to be verifiable by the network nodes. One way to achieve this, is to sign the whole image using a suitable digital signature algorithm and verify it at the receiving node. However, this strategy requires the whole image to be received prior to verification, thus wasting valuable resources in case the received image has not been properly signed. 

Another option is to sign each different page of the image separately, or even to sign each network packet independently from the others. This would enable pipelining during dissemination and save network bandwidth but could increase processing cost at a receiving node, as digital signature verification should be performed for each individual page or packet.   







Several security-related firmware update contributions are discussed in the next sections (summarised in Table~\ref{tbl:security_summary}).

\subsection{Selective 'n' Secure OTAP protocol (SenSeOP)}
The \emph{SenSeOP} OTAP protocol~\cite{aschenbruck_selective_2012} aims to secure the nodes of a wireless network from malicious and unauthorised reprogramming attempts, using an asymmetric cryptography approach based on the TinyECC library~\cite{liu_tinyecc_2008} that provides the necessary cryptographic operations for signature generation and verification. Similarly to \emph{Deluge}, \emph{SenSeOP} splits the update image into pages; however, it uses a simpler dissemination mechanism. In order to ensure the \emph{integrity} and the \emph{authenticity} of a firmware update image, it leverages ECC digital signatures using 192-bit length private keys. Moreover, using asymmetric cryptography, only the public key is stored in each node; hence, memory attacks are ineffective. The authors assume infrequent and non-regular software updates where confidentiality is not required, so no encryption is used. 

In order to save valuable resources, \emph{SenseOP} computes the digital signature of the whole firmware image and not the signatures of the packets the image is subdivided to. This restricts the signature verification at the receiving side to be performed only when the whole image is received. 

Initially, the firmware update image is hashed and encrypted using the operator's private key to produce the corresponding signature. Afterwards, the image is fragmented and the corresponding packets are transmitted through the network; broadcast, as well as unicast transmission is possible. Replay attacks are avoided through the use of a version counter combined with the destination (or broadcast) address that are included in the hash computation as well.


\subsection{Secure and DoS-Resistant Code Dissemination in Wireless Sensor Networks (Seluge)}
\emph{Seluge}~\cite{hyun_seluge_2008} is a secure extension for \emph{Deluge}, which provides integrity assurance for the firmware image and resistance against DoS attacks that specifically target firmware dissemination protocols. \emph{Seluge} provides immediate authentication of each individual packet upon receipt, defeating the DoS attacks that exploit the authentication delays the node faces when waiting to receive the rest of a page. Moreover, all advertisements and requests are authenticated using a weak authentication scheme along with a signature, since it can be efficiently verified by a node; it still takes a vast amount of time for an attacker to forge the authenticator.

The firmware image is split into fixed-size pages and each page is further divided into a number of packets. For every packet of a page $P$, a hash value is computed that is appended to the corresponding packet of page $P-1$. This is an iterative process, forming a hash tree~\cite{shoufan_fast_2010} that is used as the basis for the computation of the final signature. For the digital signatures, the ECDSA algorithm is used over the 160-bit elliptic curve \emph{secp160k1}.  

To disseminate the image, the source node first broadcasts the signature packet that serves as an advertisement for the new firmware image. Upon reception of this packet, a node cryptographically verifies it, and then uses the root of the Merkle hash tree to authenticate each hash packet of the first page. Next, since the packets of the first page carry the hash values of those of the second page, signature verification continues, and this is repeated till all pages are verified. Nevertheless, this signature scheme is vulnerable to DoS attacks as an adversary can inject bogus signature packets and force the nodes to perform energy and time-consuming operations.

To address this issue, \emph{Seluge} uses a weak authentication mechanism, called \emph{message specific puzzles}~\cite{ning_mitigating_2008}, authenticating the signature packet for a specific firmware version with a key that is cryptographically associated with the version identity number. If verification succeeds, the node will verify the puzzle solution and authenticate the source of the signature packet. This way, the node will not perform any operation for bogus signature packets. A disadvantage of \emph{Seluge} is that it increases the ROM and RAM utilisation, as a number of metadata and hash values have to be stored.

\subsection{Secure dissemination of code updates in sensor networks (Sluice)}
\emph{Sluice}~\cite{lanigan_sluice_2006} is based on \emph{Deluge} and uses \emph{hash chains} to ensure the authentication and integrity of the received firmware image, while providing pipeline support. Similarly to \emph{Deluge}, \emph{Sluice} splits each firmware image into several pages and integrates signatures and hash functions for efficient code authentication. More specifically, the hash image of each page is computed and is appended to the previous page, forming a chain of hashes. Following the concept of digital stream signing, the head of the hash chain (first page in the chain) only is signed using the private key of the firmware provider, requiring only one signature to be computed for the whole update. 

Using the digital signature, a node can verify the source of the update and can also ensure the integrity of the image, comparing  the hash found in a received page with the computed hash value of the previous one; hence, using \emph{Sluice}, there is a minimal overhead, as the only operations required are the signature validation and the computation of a few hash values. For the digital signature, the ECDSA algorithm is used with 160-bit SHA-1 hashes. 

\subsection{Securing Deluge}
In~\cite{dutta_securing_2006}, the authors propose a secure version of the \emph{Deluge} protocol that  utilises authenticated digital streams. A firmware image is transformed into a series of messages, each one containing the hash of the previous message. The  head of this hash chain is signed using RSA signatures and 64-bit SHA-1 hashes. 

In more detail, the firmware image is split into pages, where each page is further divided into a group of packets, and a hash value is computed for every such packet that is attached to the data of its previous packet. Finally, the hash value of the first packet is digitally signed producing the signature, and these two values form  the \emph{advertisement packet} of the update, the first packet broadcast when a new firmware version becomes available. If there is a receiving node that wishes a firmware update, it checks the signature and caches the hash value of the first packet. Afterwards, the normal dissemination methodology of \emph{Deluge} is followed to request the packets of the first page.

In order to ensure that packets arrive in order, the hash of each received packet is compared with the one that was stored in the last accepted one. If this comparison fails, the node requests a retransmission with a selective NACK message; hence the packets, should be received in order so that they can be accepted as legitimate packets of the firmware image; but out-of-order packets are also cached for optimisation purposes.

\subsection{Secure firmware updates using open standards}
The authors in~\cite{zandberg_2019} propose a firmware update mechanism based on open standards such as CoAP, LwM2M, SUIT, etc. The firmware server signs the firmware image and its metadata (manifest) using ECC and more specifically, the \emph{ed25519} and \emph{ECDSA}/\emph{p256r1} algorithms and elliptic curves. There is a two-process approach that gives higher flexibility: first only the metadata are transmitted, and if successfully verified by the receiving node (using a trust anchor with knowledge of firmware provider's public key), the firmware image is downloaded using CoAP block operations. No deltas are used, and during transmission, neither the metadata, nor the firmware image are encrypted.     

\subsection{Secure software update of realistic embedded devices (ASSURED)}
The authors in~\cite{asokan_2018} propose a scalable architecture for OTAP, supporting end-to-end  security. They distinguish four types of stakeholders: (i) original equipment manufacturer (OEM), (ii) firmware distributor, (iii) domain controller, and (iv) connected devices. \emph{OEM} cryptographically signs a new firmware version using ECC, based on \emph{Ed25519}, and the devices verify the signature prior to installing this new version. Firmware distributor's role is solely firmware distribution and can be a non-trusted entity, while the domain controller can set policies on the firmware update process (e.g. use of firmware deltas) that are included within a metadata structure (manifest). 

\emph{ASSURED} was built in two proof-of-concept implementations: (i) on Hydra~\cite{Eldefrawy_2017}, a hybrid (HW/SW) remote attestation design based on a micro-kernel, which offers process memory isolation and enforces access control to memory regions, and (ii) on ARM Cortex-M23 MCU that is equipped with Trustzone security extensions~\cite{pinto_2019} and is able to partition the system into two regions (secure, non-secure).

\subsection{Secure FOTA object for IoT}
In~\cite{doddapaneni_secure_2017}, the authors propose the use of a standardised approach and structure, commonly referred as \emph{objects} that can be used by IoT manufacturers. This secure object is called \emph{FOSE}. The motivation is that the packets that compose the firmware update image are usually secured by application layer security. However, the connectivity over which the transmission takes place, usually breaks and it is impossible to resume later. \emph{FOSE} ensures that no tampering of data will take place and broken connection can be resumed in a later stage. In order to be able to resume the connection, the client ACKs each received \emph{FOSE object} and the server keeps track of the counter. The data contained inside a \emph{FOSE object} are encrypted, thus providing confidentiality.

\begin{table*}[ht]
\caption{Security-enabled OTAP protocols}
\centering
\renewcommand{\arraystretch}{1.5}
\begin{tabular}{| L{0.12\textwidth} | L{0.11\textwidth} |  L{0.09\textwidth} | L{0.09\textwidth} | L{0.14\textwidth}| L{0.15\textwidth}| }
\hline
    \textbf{Protocol} & \textbf{Confidentiality} & \textbf{Integrity} & \textbf{Authenticity} & \textbf{Digital signature scheme} & \textbf{Protection against} \\
\hhline{|=|=|=|=|=|=|}
\textbf{SenSeOP}~\cite{aschenbruck_selective_2012} &
No &
Yes &
Yes &
SHA-1 \& ECC &
DoS, replay attacks \\
\hline
\textbf{Seluge}~\cite{hyun_seluge_2008} &
No &
Yes &
Yes &
ECDSA &
DoS, integrity, insider attacks \\
\hline
\textbf{Sluice}~\cite{lanigan_sluice_2006} &
No &
Yes &
Yes &
ECDSA &
Integrity, insider attacks \\
\hline
\textbf{Securing deluge}~\cite{dutta_securing_2006} &
No &
Yes &
Yes &
SHA-1, RSA &
Integrity, insider attacks \\
\hline
\textbf{Secure firmware updates using open standards}~\cite{zandberg_2019} &
No &
Yes &
Yes &
ed25519, EdDSA &
Integrity attacks \\
\hline
\textbf{ASSURED}~\cite{asokan_2018} &
Yes &
Yes &
Yes &
ed25519, EdDSA &
Integrity, insider attacks \\
\hline
\textbf{Secure FOTA object}~\cite{doddapaneni_secure_2017} &
Yes &
Yes &
Yes &
- &
Integrity attacks \\
\hline
\end{tabular}
\label{tbl:security_summary}
\end{table*}






\section{Platforms supporting firmware over-the-air programming}

In this section, we present a collection of cloud platforms that offer firmware OTAP for IoT devices. Some of them are part of a broader IoT ecosystem that may support complex application domains, while others are exclusively focused on IoT device management, the OTAP software update being part of it. All information presented here originates mainly from the documentation provided for each platform.

We focus on platforms that aim at providing robust, reliable and secure OTAP. This is guaranteed by characteristics, such as atomic software installation, easy rollback to previous software version (e.g. through A/B update that alternates two slots/partitions for loading and storing the new software), update failure management (i.e. in case of power or connectivity loss), short downtime, secure communication during software downloading, as well as authenticity and integrity verification of new software. Table~\ref{tab:platforms_summary} summarises the main aspects of the platforms under consideration.

\begin{table*}[ht]
\caption{Summary of platforms supporting Over-The-Air programming}
\centering
\renewcommand{\arraystretch}{1.5}
\begin{tabular}{|>{\bfseries}L{0.14\textwidth} || L{0.14\textwidth} | L{0.14\textwidth} | L{0.14\textwidth} | L{0.14\textwidth} | L{0.14\textwidth} | }
\hline
\textbf{Platform} & \textbf{Mender}~\cite{plat:Menderio} & \textbf{ARM Pelion}~\cite{plat:ArmPelion} & \textbf{Balena}~\cite{plat:Balena} & \textbf{Particle}~\cite{plat:Particle} & \textbf{AWS IoT - FreeRTOS}~\cite{plat:FreeRTOS} \\
\hhline{|=|=|=|=|=|=|}
Supported processors &
ARM, x86(-64) &
ARM Cortex-M, ARM Cortex-A &
ARM, x86(-64) &
ARM Cortex-M &
ARM Cortex-M, MIPS Warrior-M, Tensilica Xtensa LX6 \\
\hline
Operating system &
Embedded Linux (Yocto Project), Debian Family &
ARM MbedOS &
BalenaOS \newline (built on Yocto Project) &
Particle \newline Device OS &
Amazon FreeRTOS \\
\hline
Connectivity &
WiFi, Cellular &
WiFi, Cellular, BLE, \mbox{IEEE 802.15.4}, LoRa &
WiFi, Cellular &
WiFi, Cellular (2G, 3G, LTE), BLE, \mbox{IEEE 802.15.4} &
WiFi, BLE \\
\hline
Device authentication &
Public key, \newline JSON Web Token &
Public key, \newline X.509 Certificate &
API key & 
Public Key &
Public key, \newline X.509 Certificate \\
\hline
Communication security &
TLS & 
TLS, DTLS &
TLS &
DTLS, AES &
TLS \\
\hline
Software/firmware authenticity and integrity &
Cryptographic signatures \newline (RSA, ECDSA) &
Cryptographic signatures \newline (ECDSA) &
- &
CRC32 (non-cryptographic) &
Cryptographic signatures \newline (RSA, ECDSA) \\
\hline
Update reliability and efficiency &
\begin{minipage}[t]{0.14\textwidth}
\raggedright
\begin{itemize}[leftmargin=*]
    \setlength\itemsep{0.2em}
    \item[--] A/B update (Dual rootfs)
    \item[--] Update rollback
    \item[--] Batch update
\end{itemize}
\end{minipage}
&
\begin{minipage}[t]{0.14\textwidth}
\raggedright
\begin{itemize}[leftmargin=*]
    \setlength\itemsep{0.2em}
    \item[--] A/B update
    \item[--] Update rollback
    \item[--] Conditional update
    \item[--] Differential update
    \item[--] Continuous update
    \item[--] Batch update
\end{itemize}
\end{minipage}
& 
\begin{minipage}[t]{0.14\textwidth}
\raggedright
\begin{itemize}[leftmargin=*]
    \setlength\itemsep{0.2em}
    \item[--] A/B update (Dual rootfs)
    \item[--] Update rollback (includes boot partition)
\end{itemize}
\end{minipage}
& 
\begin{minipage}[t]{0.14\textwidth}
\raggedright
\begin{itemize}[leftmargin=*]
    \setlength\itemsep{0.2em}
    \item[--] A/B update
    \item[--] Update rollback
    \item[--] Context-aware update
    \item[--] Update on wake-up
    \item[--] Batch update
\end{itemize}
\end{minipage}
&
\begin{minipage}[t]{0.14\textwidth}
\raggedright
\begin{itemize}[leftmargin=*]
    \setlength\itemsep{0.2em}
    \item[--] A/B update
    \item[--] Update rollback
    \item[--] Context-aware update
    \item[--] Update on wake-up
    \item[--] Batch update
\end{itemize}
\end{minipage}
\\
\hline
\end{tabular}
\label{tab:platforms_summary}
\end{table*}

\subsection{Mender}
\emph{Mender}~\cite{plat:Menderio} is an open source over-the-air software update manager for embedded Linux devices, which considers security and reliability of the update process, and both application and full system update are possible. \emph{Mender} architecture is essentially built on two components: (i) \emph{Mender Management Server}, and (ii) \emph{Mender Client}. \emph{Mender Management Server} is the central point for deploying updates to IoT devices. It monitors the software that is installed on each registered device and schedules the roll-out of new releases. Devices can be organised into groups, so that batch software updates can be orchestrated. \emph{Mender Client} runs on the device and periodically polls the \emph{Mender Management Server} for monitoring reasons (e.g. status reporting), as well as for discovering pending software updates. In case software updates exist for the specific device, \emph{Mender Client} is responsible for downloading and installing it.

Obviously, a software build system is necessary for generating new device software. Mender uses \emph{Yocto Project}~\cite{plat:YoctoProject} for building the artifacts required by the target device. \emph{Yocto Project} is a Linux Foundation collaborative open source project that offers tools and processes for creating custom embedded Linux distributions. It follows a layer model for developing logically independent software pieces that can be easily customised, combined and reused, supporting its architecturally agnostic nature (it supports all major embedded architectures, such as ARM, 32-bit and 64-bit x86, PowerPC, and MIPS). Mender provides \emph{meta-mender}, a set of Yocto Project layers for embedding \emph{Mender Client} into the OS image.
Reliability of the update process is enhanced by using dual rootfs updates (new software is deployed in inactive partition that becomes active after reboot), sanity checks during first reboot and rollback to former software, if the sanity checks fail.

\emph{Mender} platform considers device authentication, software authenticity and integrity, as well as secure communication between the \emph{Mender Client} and the Mender Management Server. Initially, each device authenticates to the Management Server through an authentication set (identity attributes and public key). Subsequently, it is provided with an authentication token (JSON Web Token), by means of which each subsequent request is authenticated. Authenticity and integrity of the built software is guaranteed through cryptographic signatures that are verified at the device. Currently, the platform supports two algorithms, namely RSA with recommended key length of at least 3072 bits, and ECDSA with ECC \texttt{prime256v1} curve. Finally, communication between devices and back-end is secured through Transport Layer Security (TLS).

It is noted that although \emph{Mender} is not a general purpose IoT device management platform, as it is solely focused on managing and orchestrating the software updates, it has been successfully integrated into other major IoT platforms, such as Google Cloud IoT Core and Microsoft Azure IoT.

\subsection{ARM Pelion}
ARM provides a full-stack IoT solution, spanning from embedded devices to IoT cloud services. \emph{ARM Pelion}~\cite{plat:ArmPelion} IoT platform is a suite of management services that focuses on three core IoT components, namely connectivity, device, and data management, for devices running the ARM Mbed OS or Linux/Mbed Linux OS. Several different processor architectures are supported, from simple Cortex-M microcontrollers to powerful Cortex-A systems, as well as both IP and non-IP based communication protocols (e.g. LoRa, BLE), the later with the support of an appropriate gateway that employs the necessary protocol translation.

Communication between the IoT devices and the Management Server is based on the Open Mobile Alliance \emph{Lightweight Machine-to-Machine} (LwM2M) application protocol that is used in combination with the \emph{Constrained Application Protocol} (CoAP), over UDP or TCP transports. Transport layer security is provided by Datagram Transport Layer Security (DTLS) or TLS protocol, respectively. The LwM2M protocol provides a simple, yet very efficient, data model for: (i) device bootstrapping, (ii) discovery and registration, (iii) device management and service enablement, and (iv) information reporting. It is noted that the LwM2M data model provides native support for firmware update process, by defining the necessary resources the device needs to expose in order to receive the firmware binary.

Over-the-air firmware updates are performed in the form of \emph{campaigns} that apply either to a single device or fleet of devices. A typical full firmware image contains the OS, the \emph{Device Management Update Client} (responsible for managing the update process on the device part) and the user application. Efficiency of the update process, especially for low-rate and low-power IoT devices (e.g. NB-IoT), is enhanced by delta updates (binary patches for constructing new firmware binary from the existing one). Reliability mechanisms, include \emph{conditional updates} (a device accepts updates based on pre-defined conditions, e.g. minimum battery level), \emph{sanity checks} on received firmware (bootloader verifies firmware integrity, by calculating its hash and comparing it with the one received as firmware metadata) and \emph{rollback support} through dual partitions (active partition / candidate partition). Each firmware image is accompanied by a piece of information, named as \emph{manifest}. The manifest is essentially firmware metadata that encodes information on firmware authenticity, integrity, device compatibility and update logistics (update time scheduling, binary storage options etc.).

\emph{ARM Pelion Device Management} uses Public Key Infrastructure (PKI)-based security, and relies on X.509 certificates and public-key encryption for server and device authentication. Authenticity and cryptographic integrity of a firmware binary and its corresponding manifest is achieved by ECDSA signatures (based on ECC \texttt{secp256r1} curve).

In addition to the device management services, ARM provides an online compiler, named as \emph{Mbed Compiler}, for developing and building device firmware through a web SDK. The user can download the produced binary either locally or directly use Pelion services for performing a firmware OTA update campaign to registered devices. Mbed Compiler integrates features, such as source code version control and collaboration tools for multi-author projects. Finally, it hosts a large database of free user-created libraries that can be easily imported into any application.

\subsection{Balena}
\emph{Balena}~\cite{plat:Balena} offers a complete set of tools for building, deploying, and managing fleets of connected embedded Linux IoT devices. It builds on Linux containers technology for easily deploying updates on applications or even the entire host OS running on an IoT device.

The main software components of \emph{Balena} ecosystem are described in the following. IoT devices run \emph{Balena} OS, a Yocto Project Linux-based OS, packaged with \emph{balenaEngine}, a lightweight Docker-compatible container engine that manages containers executing \emph{Balena} services or user applications. On the cloud side, the \emph{BalenaCloud} platform is responsible for device and communication management, as well as source code version control (though \texttt{git} repositories), software build, storage and device update. A set of CLI tools is also provided for local development, local or remote software building and device software deployment.

A device authenticates to BalenaCloud by using API keys. During provisioning it receives a provisioning API key that is used for authentication after first boot-up and registration to the back-end. Then, a new API key is generated and provided to the device, which uses it for subsequent authentication. Device control and image download during updates is performed over a Virtual Private Network (VPN), so that all traffic is secured with TLS.

Reliability of the update process is based on pre-update sanity checks in terms of software-device compatibility and availability of the image in the container registry, double root partition approach (active/inactive partition) and rollback support, in case boot partition changes fail (remember that it is possible to update the entire host operating system, not only a user application that is executed on the device).

\subsection{Particle}
\emph{Particle}~\cite{plat:Particle} offers a full-stack IoT solution that includes different hardware platforms, various connectivity options and cloud services for device fleet management, over-the-air updates and device health monitoring. Microprocessor architectures supported are of ARM Cortex-M series, while connectivity options include WiFi, Cellular (2G, 3G, LTE), BLE and IEEE 802.15.4. Devices run the \emph{Particle} \emph{Device OS} operating system that provides the necessary hardware abstraction for easy application development and enables connectivity and management functionalities. On the other end, the \emph{Device Cloud} is the cloud back-end that provides device management and monitoring, through real-time event and data logging, firmware roll-outs and over-the-air updates, as well as integration with other major IoT platforms (e.g. Google Cloud Platform, Azure IoT Hub) and data publishing through Webhooks.

Mutual authentication between Device Cloud and the devices is based on RSA public/private key pairs. Communication is encrypted by using DTLS over UDP or AES over TCP (depending on the device capabilities), and CoAP is used for data collection and device management operations, including over-the-air firmware update.

Reliability and resiliency of the firmware update process is supported by features, such as \emph{atomic updates}, according to which only a fully received and verified firmware is executed, \emph{automatic rollbacks}, in case of failure during firmware transfer, \emph{context-aware updates} based on device operational status (e.g. update is postponed for a later time, if the device currently performs a critical task), \emph{update on wake-up}, for currently sleeping devices and \emph{batch updates}. It is noted that updates are possible both at an application and Device OS level, and can be initiated either through provided CLI tools, Device Cloud Web UI or Device Cloud REST API.

The platform offers additionally a Web IDE, named as \emph{Pacticle Build} and a database of libraries for developing and building device firmware in the cloud, which can be further released to registered device fleets.

\subsection{AWS IoT - FreeRTOS}
\emph{FreeRTOS}~\cite{plat:FreeRTOS} is a well-known preemptive real-time operating system for embedded devices that employs a priority-based dynamic scheduler and a multi-threading programming model. It is open-source and has been successfully ported to a large number of microcontroller/microprocessor platforms, including ARM variants (ARM7, ARM9, Cortex-M and Cortex-A Series), Atmel AVR, MSP430, Espressif ESP32 etc. An extension of FreeRTOS, provided by Amazon, includes necessary libraries for the secure and reliable connection of IoT devices with \emph{Amazon Web Services IoT} (AWS-IoT)~\cite{plat:AWSIoT} cloud platform, for device management, data collection and application development. IoT devices communicate with the cloud back-end through the Message Queue Telemetry Transport (MQTT) protocol, either directly or over Websockets. In addition, an HTTP REST endpoint exists for data publishing. Mutual authentication is achieved by using PKI X.509 certificates and the TLS protocol is used for securing the communication channel between IoT devices and AWS IoT back-end.

Over-the-air firmware update is supported for devices running Amazon FreeRTOS. The \emph{OTA Update Manager} service, which is part of the AWS IoT backend, is responsible for notifying the device on existing updates, orchestrating the update process and maintaining the update log. On the device side, the \emph{OTA Agent} module manages the notification, downloading (over secure MQTT or HTTP) and verification of the firmware updates. In order to ensure the authenticity and integrity of the firmware, the OTA Update Manager cryptographically signs the firmware image before deployment (both RSA and ECDSA signatures are supported). The OTA Agent verifies the signature before applying any new update.

In order to improve reliability and efficiency of the update process, the AWS IoT for FreeRTOS provides support for \emph{batch updates} (device group or entire fleet), \emph{continuous updates} (so that new firmware is deployed to devices as they are added to groups, are reset or re-provisioned), as well as \emph{update rollback}.

\section{Conclusion}
The proliferation of massive IoT networks has been remarkable in the last years. These ubiquitous smart object-enabled networks, which may operate for several years in variable conditions, are used for supporting complex applications in several domains, such as smart cities, healthcare, industrial automation, etc. Throughout their extended lifetime, the nodes forming the IoT networks need to be re-programmed, so that new features are added, software bugs or security vulnerabilities are resolved and their functionality is re-purposed. The large scale of IoT networks and the usual installation of IoT nodes in locations with difficult or no physical access, mandates the use of OTAP solutions for the efficient update of firmware running on IoT nodes.

In this paper, we present an overview of OTAP techniques that can be applied to IoT networks. We highlight the main challenges and limitations stemming from the resource-constraint and heterogeneous nature of IoT nodes and analyse the essential stages of firmware update process, along with different approaches and techniques that implement them. In addition, we discuss schemes that focus on securing the OTAP process by encrypting the transmitted firmware and/or providing firmware authenticity and integrity-preserving mechanisms. Finally, we present a collection of state-of-the art of commercial and open-source platforms that integrate secure and reliable OTAP.

\ifCLASSOPTIONcompsoc
  \section*{Acknowledgments}
\else
  \section*{Acknowledgment}
\fi

{This research has been financed by the European Union and Greek national funds through the Operational Program Competitiveness, Entrepreneurship and Innovation, under the call RESEARCH – CREATE – INNOVATE (project code: T1EDK-03389).}

\ifCLASSOPTIONcaptionsoff
  \newpage
\fi

\bibliographystyle{IEEEtran}
\bibliography{all_bib}

\begin{thebibliography}{100}
\providecommand{\url}[1]{#1}
\csname url@samestyle\endcsname
\providecommand{\newblock}{\relax}
\providecommand{\bibinfo}[2]{#2}
\providecommand{\BIBentrySTDinterwordspacing}{\spaceskip=0pt\relax}
\providecommand{\BIBentryALTinterwordstretchfactor}{4}
\providecommand{\BIBentryALTinterwordspacing}{\spaceskip=\fontdimen2\font plus
\BIBentryALTinterwordstretchfactor\fontdimen3\font minus
  \fontdimen4\font\relax}
\providecommand{\BIBforeignlanguage}[2]{{%
\expandafter\ifx\csname l@#1\endcsname\relax
\typeout{** WARNING: IEEEtran.bst: No hyphenation pattern has been}%
\typeout{** loaded for the language `#1'. Using the pattern for}%
\typeout{** the default language instead.}%
\else
\language=\csname l@#1\endcsname
\fi
#2}}
\providecommand{\BIBdecl}{\relax}
\BIBdecl

\bibitem{Lin_17}
J.~Lin, W.~Yu, N.~Zhang, X.~Yang, H.~Zhang, and W.~Zhao, ``A survey on
  {I}nternet of {T}hings: {A}rchitecture, enabling technologies, security and
  privacy, and applications,'' \emph{IEEE Internet of Things Journal}, vol.~4,
  pp. 1125--1142, 2017.

\bibitem{dunkels_run-time_2006}
A.~Dunkels, N.~Finne, J.~Eriksson, and T.~Voigt, ``Run-time dynamic linking for
  reprogramming wireless sensor networks,'' in \emph{Proc. of the 4th
  international conference on {Embedded} networked sensor systems - {SenSys}
  '06}.\hskip 1em plus 0.5em minus 0.4em\relax ACM Press, 2006, p.~15.

\bibitem{shi_survey_2011}
J.~Shi, J.~Wan, H.~Yan, and H.~Suo, ``A survey of {Cyber}-{Physical}
  {Systems},'' in \emph{Proc. of the {International} {Conference} on {Wireless}
  {Communications} and {Signal} {Processing} ({WCSP})}, 2011, pp. 1--6.

\bibitem{wilson_sensors_1999}
C.~Wilson, ``Sensors in medicine,'' \emph{The Western journal of medicine},
  1999.

\bibitem{crossbow_technology_inc_2003_mote_nodate}
C.~Technology, ``Mote {In}-{Network} {Programming} {User} {Reference} {Version}
  20030315. {Crossbow} {Technology}, {Inc}.''

\bibitem{stolikj_efficient_2013}
M.~Stolikj, P.~J. Cuijpers, and J.~Lukkien, ``Efficient reprogramming of
  wireless sensor networks using incremental updates,'' in \emph{2013 {IEEE}
  {International} {Conference} on {Pervasive} {Computing} and {Communications}
  {Workshops} ({PERCOM} {Workshops})}, 2013, pp. 584--589.

\bibitem{liu_implementing_2004}
T.~Liu, C.~Sadler, P.~Zhang, and M.~Martonosi, ``Implementing software on
  resource-constrained mobile sensors: experiences with {Impala} and
  {ZebraNet},'' in \emph{Proceedings of the 2nd international conference on
  {Mobile} systems, applications, and services - {MobiSYS} '04}.\hskip 1em plus
  0.5em minus 0.4em\relax ACM Press, 2004, p. 256.

\bibitem{noauthor_rfc_nodate}
\BIBentryALTinterwordspacing
``{RFC} 7228 - {Terminology} for {Constrained}-{Node} {Networks}.'' [Online].
  Available: \url{https://datatracker.ietf.org/doc/rfc7228/}
\BIBentrySTDinterwordspacing

\bibitem{baccelli_scripting_2018}
E.~Baccelli, J.~Doerr, S.~Kikuchi, F.~Padilla, K.~Schleiser, and I.~Thomas,
  ``Scripting {Over}-{The}-{Air}: {Towards} {Containers} on {Low}-end {Devices}
  in the {Internet} of {Things},'' in \emph{2018 {IEEE} {International}
  {Conference} on {Pervasive} {Computing} and {Communications} {Workshops}
  ({PerCom} {Workshops})}.\hskip 1em plus 0.5em minus 0.4em\relax IEEE, 2018,
  pp. 504--507.

\bibitem{qiang_wang_reprogramming_2006}
{Qiang W.}, {Yaoyao Z.}, and {L. Cheng}, ``Reprogramming wireless sensor
  networks: challenges and approaches,'' \emph{IEEE Network}, vol.~20, no.~3,
  pp. 48--55, 2006.

\bibitem{brown_software_2013}
S.~Brown and C.~Sreenan, ``Software {Updating} in {Wireless} {Sensor}
  {Networks}: {A} {Survey} and {Lacunae},'' \emph{Journal of Sensor and
  Actuator Networks}, pp. 717--760, 2013.

\bibitem{zandberg_secure_2019}
K.~Zandberg, K.~Schleiser, F.~Acosta, H.~Tschofenig, and E.~Baccelli, ``Secure
  {Firmware} {Updates} for {Constrained} {IoT} {Devices} {Using} {Open}
  {Standards}: {A} {Reality} {Check},'' \emph{IEEE Access}, 2019.

\bibitem{wang_survey_2006}
Y.~Wang, G.~Attebury, and B.~Ramamurthy, ``A survey of security issues in
  wireless sensor networks,'' \emph{IEEE Communications Surveys Tutorials}, pp.
  2--23, 2006.

\bibitem{chen_survey_2016}
D.~Chen, D.~He, and F.~Ahmad, ``A survey of reprogramming security in wireless
  sensor network,'' \emph{VFAST Transactions on Software Engineering}, 2016.

\bibitem{Bauwens_20}
J.~Bauwens, P.~Ruckebusch, S.~Giannoulis, I.~Moerman, and E.~D. Poorter,
  ``Over-the-air software updates in the internet of things: An overview of key
  principles,'' \emph{IEEE Communications Magazine}, pp. 35--41, 2020.

\bibitem{sanvido_nand_2008}
M.~Sanvido, F.~Chu, A.~Kulkarni, and R.~Selinger, ``nand {Flash} {Memory} and
  {Its} {Role} in {Storage} {Architectures},'' \emph{Proc. of the IEEE},
  vol.~96, pp. 1864--1874, 2008.

\bibitem{yiu_introduction_2015}
J.~Yiu, ``Introduction to {Embedded} {Software} {Development},'' in \emph{The
  {Definitive} {Guide} to {Arm}® {Cortex}®-{M0} and {Cortex}-{M0}+
  {Processors}}.\hskip 1em plus 0.5em minus 0.4em\relax Elsevier, 2015, pp.
  55--86.

\bibitem{panta_hermes_2009}
R.~Panta and S.~Bagchi, ``Hermes: {Fast} and {Energy} {Efficient} {Incremental}
  {Code} {Updates} for {Wireless} {Sensor} {Networks},'' in \emph{{IEEE}
  {INFOCOM} 2009 - {The} 28th {Conference} on {Computer}
  {Communications}}.\hskip 1em plus 0.5em minus 0.4em\relax IEEE, 2009, pp.
  639--647.

\bibitem{panta_zephyr_2009}
R.~Panta, S.~Bagchi, and S.~Midkiff, ``Zephyr: efficient incremental
  reprogramming of sensor nodes using function call indirections and difference
  computation,'' in \emph{USENIX}, 2009.

\bibitem{shafi_no-reboot_2012}
N.~Shafi, K.~Ali, and H.~Hassanein, ``No-reboot and zero-flash over-the-air
  programming for {Wireless} {Sensor} {Networks},'' in \emph{Proc. of the 9th
  {Annual} {IEEE} {Communications} {Society} {Conference} on {Sensor}, {Mesh}
  and {Ad} {Hoc} {Communications} and {Networks} ({SECON})}, 2012, pp.
  371--379.

\bibitem{frisch_over_2017}
D.~Frisch, S.~Reißmann, and C.~Pape, ``An {Over} the {Air} {Update}
  {Mechanism} for {ESP8266} {Microcontrollers},'' 2017.

\bibitem{farooq_operating_2011}
M.~Farooq and T.~Kunz, ``Operating {Systems} for {Wireless} {Sensor}
  {Networks}: {A} {Survey},'' \emph{Sensors}, pp. 5900--5930, 2011.

\bibitem{nisha_rsa_2017}
S.~Nisha and M.~Farik, ``{RSA} {Public} {Key} {Cryptography} {Algorithm} –
  {A} {Review},'' \emph{International Journal of Scientific \& Technology
  Research}, pp. 187--191, 2017.

\bibitem{dsouza_2017}
F.~D'Souza and D.~Panchal, ``Advanced encryption standard (aes) security
  enhancement using hybrid approach,'' in \emph{Proceedings of the
  International Conference on Computing, Communication and Automation (ICCCA)},
  2017, pp. 647--652.

\bibitem{mughal_2018}
M.~Mughal, X.~Luo, A.~Ullah, S.~Ullah, and Z.~Mahmood, ``A lightweight digital
  signature based security scheme for human-centered internet of things,''
  \emph{IEEE Access}, pp. 31\,630--31\,643, 2018.

\bibitem{Banerjee_2019}
U.~Banerjee, A.~Wright, C.~Juvekar, M.~Arvind, and A.~Chandrakasan, ``An
  energy-efficient reconfigurable dtls cryptographic engine for securing
  internet-of-things applications,'' \emph{IEEE Journal of Solid-state
  circuits}, vol.~54, pp. 2339--2352, 2019.

\bibitem{tal_white_2002}
A.~Tal, ``White {Paper} {Two} {Flash} {Technologies} {Compared} : {NOR} vs
  {NAND},'' 2002.

\bibitem{ondrej_effective_nodate}
\BIBentryALTinterwordspacing
O.~Kachman, ``Effective multiplatform firmware update process for embedded
  low-power devices,'' 2018. [Online]. Available:
  \url{http://acmbulletin.fiit.stuba.sk/vol11num1/kachman2019.pdf}
\BIBentrySTDinterwordspacing

\bibitem{dong_elon_2010}
W.~Dong, Y.~Liu, X.~Wu, L.~Gu, and C.~Chen, ``Elon: enabling efficient and
  long-term reprogramming for wireless sensor networks,'' in \emph{Proc. of the
  {ACM} {SIGMETRICS} international conference on {Measurement} and modeling of
  computer systems - {SIGMETRICS} '10}.\hskip 1em plus 0.5em minus 0.4em\relax
  ACM Press, 2010, p.~49.

\bibitem{kulkarni_mnp_2005}
S.~Kulkarni and L.~Wang, ``{MNP}: {Multihop} {Network} {Reprogramming}
  {Service} for {Sensor} {Networks},'' in \emph{25th {IEEE} {International}
  {Conference} on {Distributed} {Computing} {Systems} ({ICDCS}'05)}, 2005, pp.
  7--16.

\bibitem{stathopoulos_remote_2004}
T.~Stathopoulos, J.~Heidemann, and D.~Estrin, ``A {Remote} {Code} {Update}
  {Mechanism} for {Wireless} {Sensor} {Networks},'' Tech. Rep., 2004.

\bibitem{ersue_management_2015}
M.~Ersue, D.~Romascanu, J.~Schoenwaelder, and A.~Sehgal, ``Management of
  {Networks} with {Constrained} {Devices}: {Use} {Cases}, rfc7548,'' Tech.
  Rep., 2015.

\bibitem{reijers_efficient_2003}
N.~Reijers and K.~Langendoen, ``Efficient {Code} {Distribution} in {Wireless}
  {Sensor} {Networks},'' in \emph{Proc. of the 2nd {ACM} {International}
  {Conference} on {Wireless} {Sensor} {Networks} and {Applications}}.\hskip 1em
  plus 0.5em minus 0.4em\relax Association for Computing Machinery, 2003, pp.
  60--67.

\bibitem{pallister_high-level_2014}
J.~Pallister, K.~Eder, S.~Hollis, and J.~Bennett, ``A {High}-{Level} {Model} of
  {Embedded} {Flash} {Energy} {Consumption},'' in \emph{Proc. of the 2014
  {International} {Conference} on {Compilers}, {Architecture} and {Synthesis}
  for {Embedded} {Systems}}.\hskip 1em plus 0.5em minus 0.4em\relax Association
  for Computing Machinery, 2014.

\bibitem{dong_r3_2013}
W.~Dong, B.~Mo, C.~Huang, Y.~Liu, and C.~Chen, ``R3: {Optimizing} relocatable
  code for efficient reprogramming in networked embedded systems,'' in
  \emph{Proc. of the IEEE INFOCOM}, 2013, pp. 315--319.

\bibitem{dong_r2_2013}
W.~Dong, Y.~Liu, C.~Chen, J.~Bu, C.~Huang, and Z.~Zhao, ``R2: {Incremental}
  {Reprogramming} {Using} {Relocatable} {Code} in {Networked} {Embedded}
  {Systems},'' \emph{IEEE Transactions on Computers}, vol.~62, pp. 1837--1849,
  2013.

\bibitem{zhang_live_2016}
C.~Zhang, W.~Ahn, Y.~Zhang, and B.~R. Childers, ``Live code update for {IoT}
  devices in energy harvesting environments,'' in \emph{Proc. of the 5th
  {Non}-{Volatile} {Memory} {Systems} and {Applications} {Symposium}
  ({NVMSA})}.\hskip 1em plus 0.5em minus 0.4em\relax IEEE, 2016, pp. 1--6.

\bibitem{hahm_operating_2016}
O.~Hahm, E.~Baccelli, H.~Petersen, and N.~Tsiftes, ``Operating {Systems} for
  {Low}-{End} {Devices} in the {Internet} of {Things}: {A} {Survey},''
  \emph{IEEE Internet of Things Journal}, pp. 720--734, 2016.

\bibitem{aschenbruck_selective_2012}
N.~Aschenbruck, J.~Bauer, J.~Bieling, A.~Bothe, and M.~Schwamborn, ``Selective
  and {Secure} {Over}-{The}-{Air} {Programming} for {Wireless} {Sensor}
  {Networks},'' in \emph{2012 21st {International} {Conference} on {Computer}
  {Communications} and {Networks} ({ICCCN})}, 2012, pp. 1--6.

\bibitem{zhu_detection_2018}
H.~Zhu, Z.~Zhang, J.~Du, S.~Luo, and Y.~Xin, ``Detection of selective
  forwarding attacks based on adaptive learning automata and communication
  quality in wireless sensor networks,'' \emph{International Journal of
  Distributed Sensor Networks}, vol.~14, 2018.

\bibitem{frimpong_iot-cryptodiet_2020}
E.~Frimpong and A.~Michalas, ``{IoT}-{CryptoDiet}: {Implementing} a
  {Lightweight} {Cryptographic} {Library} based on {ECDH} and {ECDSA} for the
  {Development} of {Secure} and {Privacy}-{Preserving} {Protocols} in
  {Contiki}-{NG},'' 2020.

\bibitem{lakkundi_lightweight_2014}
V.~Lakkundi and K.~Singh, ``Lightweight {DTLS} implementation in {CoAP}-based
  {Internet} of {Things},'' in \emph{Proceedings of ADCOM}, 2014.

\bibitem{TragosBC_17}
E.~Tragos, A.~Fragkiadakis, V.~Angelakis, and H.~Pohls, ``Designing secure iot
  architectures for smart city applications,'' in \emph{Designing, Developing,
  and Facilitating Smart Cities: Urban Design and IoT Solutions}.\hskip 1em
  plus 0.5em minus 0.4em\relax Springer, 2017.

\bibitem{Fragkiadakis_16}
A.~Fragkiadakis and E.~Tragos, ``A trust-based scheme employing evidence
  reasoning for iot architectures,'' in \emph{IEEE 3rd World Forum on Internet
  of Things (WF-IoT)}, 2016, pp. 559--564.

\bibitem{Lili_18}
L.~Li, Z.~Xi, Y.~Zhu, and S.~Wang, ``Improvement and implementation of rpl
  routing protocol in wireless sensor networks,'' in \emph{WiCOM}, 2018.

\bibitem{Szewczyk2004LessonsFA}
R.~Szewczyk, J.~Polastre, A.~Mainwaring, and D.~Culler, ``Lessons from a sensor
  network expedition,'' in \emph{EWSN}, 2004.

\bibitem{Antonakakis_17}
M.~Antonakakis, T.~April, M.~Bailey, M.~Bernhard, E.~Bursztein, J.~Cochran,
  Z.~Durumeric, J.~Halderman, L.~Invernizzi, M.~Kallitsis, D.~Kumar, C.~Lever,
  Z.~Ma, J.~Mason, D.~Menscher, C.~Seaman, N.~Sullivan, and K.~T. an~Y.~Zhou,
  ``Understanding the mirai botnet,'' in \emph{26th USENIX Security Symposium},
  2017, pp. 1093--1110.

\bibitem{angrishi_turning_2017}
\BIBentryALTinterwordspacing
A.~Kishore, ``Turning internet of things (iot) into internet of vulnerabilities
  (iov) : Iot botnets,'' 2017. [Online]. Available:
  \url{https://arxiv.org/abs/1702.03681v1}
\BIBentrySTDinterwordspacing

\bibitem{ronen_iot_2017}
E.~Ronen, A.~Shamir, A.~Weingarten, and C.~OFlynn, ``{IoT} {Goes} {Nuclear}:
  {Creating} a {ZigBee} {Chain} {Reaction},'' in \emph{Proceedings of IEEE
  Symposium on Security and Privacy}, 2017.

\bibitem{kachman_configurable_2016}
\BIBentryALTinterwordspacing
O.~Kachman, ``Configurable {Reprogramming} {Scheme} for {Over}-{theAir}
  {Updates} in {Networked} {Embedded} {Systems},'' 2016. [Online]. Available:
  \url{http://www.fit.vutbr.cz/events/pad2016/download/sbornik/11-Kachman.pdf}
\BIBentrySTDinterwordspacing

\bibitem{park_non-invasive_2014}
H.~Park, J.~Jeong, and P.~Mah, ``Non-invasive rapid and efficient firmware
  update for wireless sensor networks,'' in \emph{Proc. of the 2014 {ACM}
  {International} {Joint} {Conference} on {Pervasive} and {Ubiquitous}
  {Computing} {Adjunct} {Publication} - {UbiComp} '14 {Adjunct}}.\hskip 1em
  plus 0.5em minus 0.4em\relax ACM Press, 2014, pp. 147--150.

\bibitem{noauthor_map_nodate}
\BIBentryALTinterwordspacing
``Map files {GNU}.'' [Online]. Available:
  \url{https://ftp.gnu.org/old-gnu/Manuals/ld-2.9.1/html_node/ld_3.html}
\BIBentrySTDinterwordspacing

\bibitem{adly_over--air_2010}
I.~Adly, H.~Ragai, A.~El-Hennawy, and K.~Shehata, ``Over-{The}-{Air}
  {Programming} of {PSoC} {Sensor} {Interface} in {Wireless} {Sensor}
  {Networks},'' in \emph{Proc. of the {Mediterranean} {Electrotechnical}
  {Conference} - {MELECON}}, 2010, pp. 997 -- 1002.

\bibitem{parthasarathy_over_2010}
R.~Parthasarathy, N.~Peterson, W.~Song, A.~Hurson, and B.~Shirazi, ``Over the
  {Air} {Programming} on {Imote2}-{Based} {Sensor} {Networks},'' in \emph{2010
  43rd {Hawaii} {International} {Conference} on {System} {Sciences}}, 2010, pp.
  1--9.

\bibitem{noauthor_tmote_2006}
\BIBentryALTinterwordspacing
``Tmote {Sky},'' 2006. [Online]. Available:
  \url{https://insense.cs.st-andrews.ac.uk/files/2013/04/tmote-sky-datasheet.pdf}
\BIBentrySTDinterwordspacing

\bibitem{noauthor_micaz_nodate}
\BIBentryALTinterwordspacing
\emph{{MICAz}}. [Online]. Available:
  \url{http://www.openautomation.net/uploadsproductos/micaz_datasheet.pdf}
\BIBentrySTDinterwordspacing

\bibitem{koshy_remote_2005}
J.~Koshy and R.~Pandey, ``Remote incremental linking for energy-efficient
  reprogramming of sensor networks,'' in \emph{Proc. of the {Second} {European}
  {Workshop} on {Wireless} {Sensor} {Networks}, 2005.}\hskip 1em plus 0.5em
  minus 0.4em\relax Istanbul, Turkey: IEEE, 2005, pp. 354--365.

\bibitem{han_sos_2005}
C.~Han, R.~Kumar, R.~Shea, E.~Kohler, and M.~Srivastava, ``{SOS} -{A} {Dynamic}
  operating system for {Sensor} {Networks},'' in \emph{Proc. of MobiSys}, 2005.

\bibitem{dunkels_contiki_2004}
A.~Dunkels, B.~Gronvall, and T.~Voigt, ``Contiki - a lightweight and flexible
  operating system for tiny networked sensors,'' in \emph{29th {Annual} {IEEE}
  {International} {Conference} on {Local} {Computer} {Networks}}, 2004, pp.
  455--462.

\bibitem{hu_reprogramming_2009}
J.~Hu, C.~J. Xue, Y.~He, and E.~H.-M. Sha, ``Reprogramming with {Minimal}
  {Transferred} {Data} on {Wireless} {Sensor} {Network},'' in \emph{Proc. of
  the 6th {International} {Conference} on {Mobile} {Adhoc} and {Sensor}
  {Systems}}.\hskip 1em plus 0.5em minus 0.4em\relax Macau, China: IEEE, 2009,
  pp. 160--167.

\bibitem{jeong_node-level_2003}
J.~Jeong, ``Node-level {Representation} and {System} {Support} for {Network}
  {Programming},'' 2003.

\bibitem{tridgell_efficient_1999}
A.~Tridgell, ``Efficient algorithms for sorting and synchronization,'' Ph.D.
  dissertation, The Australian National University, 1999.

\bibitem{korn_vcdiff_2002}
D.~Korn, J.~MacDonald, J.~Mogul, and K.~Vo, \emph{The {VCDIFF} {Generic}
  {Differencing} and {Compression} {Data} {Format}}.\hskip 1em plus 0.5em minus
  0.4em\relax RFC Editor, 2002, published: RFC 3284.

\bibitem{mazumder_efficient_2013}
B.~Mazumder and J.~O. Hallstrom, ``An efficient code update solution for
  wireless sensor network reprogramming,'' in \emph{Proc. of the 2013
  {International} {Conference} on {Embedded} {Software} ({EMSOFT})}, 2013, pp.
  1--10.

\bibitem{mo_efficient_2012}
B.~Mo, W.~Dong, C.~Chen, J.~Bu, and Q.~Wang, ``An efficient differencing
  algorithm based on suffix array for reprogramming wireless sensor networks,''
  in \emph{Proc. of the 2012 {IEEE} {International} {Conference} on
  {Communications} ({ICC})}, 2012, pp. 773--777.

\bibitem{kachman_optimized_2016}
O.~Kachman and M.~Balaz, ``Optimized differencing algorithm for firmware
  updates of low-power devices,'' in \emph{2016 {IEEE} 19th {International}
  {Symposium} on {Design} and {Diagnostics} of {Electronic} {Circuits} \&
  {Systems} ({DDECS})}.\hskip 1em plus 0.5em minus 0.4em\relax IEEE, 2016, pp.
  1--4.

\bibitem{jaein_jeong_incremental_2004}
J.~J. and D.~Culler, ``Incremental network programming for wireless sensors,''
  in \emph{Proc. of the {First} {Annual} {IEEE} {Communications} {Society}
  {Conference} on {Sensor} and {Ad} {Hoc} {Communications} and {Networks}.},
  2004, pp. 25--33.

\bibitem{hirschberg_linear_1975}
D.~Hirschberg, ``A {Linear} {Space} {Algorithm} for {Computing} {Maximal}
  {Common} {Subsequences},'' \emph{Commun. ACM}, vol.~18, pp. 341--343, 1975.

\bibitem{dementiev_better_2008}
R.~Dementiev, J.~Kärkkäinen, J.~Mehnert, and P.~Sanders, ``Better external
  memory suffix array construction,'' \emph{Journal of Experimental
  Algorithmics}, vol.~12, 2008.

\bibitem{lehniger_impact_2019}
K.~Lehniger and S.~Weidling, ``The {Impact} of {Diverse} {Execution}
  {Strategies} on {Incremental} {Code} {Updates} for {Wireless} {Sensor}
  {Networks}:,'' in \emph{Proceedings of the 8th {International} {Conference}
  on {Sensor} {Networks}}.\hskip 1em plus 0.5em minus 0.4em\relax SCITEPRESS -
  Science and Technology Publications, 2019, pp. 30--39.

\bibitem{stann_rmst_2003}
F.~Stann and J.~Heidemann, ``{RMST}: reliable data transport in sensor
  networks,'' in \emph{Proceedings of the {First} {IEEE} {International}
  {Workshop} on {Sensor} {Network} {Protocols} and {Applications}, 2003.},
  2003, pp. 102--112.

\bibitem{kulkarni_infuse_2006}
S.~Kulkarni and M.~Arumugam, ``Infuse: {A} {TDMA} {Based} {Data}
  {Dissemination} {Protocol} for {Sensor} {Networks},'' \emph{International
  Journal of Distributed Sensor Networks}, vol.~2, pp. 55--78, 2006.

\bibitem{naik_sprinkler_2005}
V.~Naik, A.~Arora, P.~Sinha, and H.~Z., ``Sprinkler: a reliable and energy
  efficient data dissemination service for wireless embedded devices,'' in
  \emph{26th {IEEE} {International} {Real}-{Time} {Systems} {Symposium}
  ({RTSS}'05)}, 2005, pp. 10 pp.--286.

\bibitem{tseng_broadcast_2002}
Y.~Tseng, S.~Ni, Y.~Chen, and J.~Sheu, ``The {Broadcast} {Storm} {Problem} in a
  {Mobile} {Ad} {Hoc} {Network},'' \emph{Wireless Networks}, vol.~8, pp.
  153--167, 2002.

\bibitem{alagar_reliable_1995}
S.~Alagar, S.~Venkatesan, and J.~Cleveland, ``Reliable broadcast in mobile
  wireless networks,'' in \emph{Proceedings of {MILCOM} '95}, vol.~1, 1995, pp.
  236--240 vol.1.

\bibitem{chlipala_deluge_2004}
A.~Chlipala, J.~Hui, and G.~Tolle, ``Deluge: {Data} dissemination for network
  reprogramming at scale,'' in \emph{Class project, Berkeley, University of
  California}, 2004.

\bibitem{hagedorn_rateless_2008}
A.~Hagedorn, D.~Starobinski, and A.~Trachtenberg, ``Rateless {Deluge}:
  {Over}-the-{Air} {Programming} of {Wireless} {Sensor} {Networks} {Using}
  {Random} {Linear} {Codes},'' in \emph{2008 {International} {Conference} on
  {Information} {Processing} in {Sensor} {Networks} (ipsn 2008)}, 2008, pp.
  457--466.

\bibitem{dong_efficient_2014}
C.~Dong and F.~Yu, ``An efficient network reprogramming protocol for wireless
  sensor networks,'' \emph{Computer Communications}, vol.~55, 2014.

\bibitem{levis_tossim_2003}
P.~Levis, N.~Lee, M.~Welsh, and D.~Culler, ``{TOSSIM}: {Accurate} and
  {Scalable} {Simulation} of {Entire} {TinyOS} {Applications},'' in
  \emph{Proceedings of the 1st international conference on Embedded networked
  sensor systems}, 2003.

\bibitem{elson_emstar_2003}
J.~Elson, S.~Bien, N.~Busek, V.~Bychkovskiy, A.~Cerpa, D.~Ganesan, L.~Girod,
  B.~Greenstein, T.~Schoellhammer, T.~Stathopoulos, and D.~Estrin, ``{EmStar}:
  {An} {Environment} for {Developing} {Wireless} {Embedded} {Systems}
  {Software},'' Tech. Rep., 2003.

\bibitem{levis_trickle_2004}
P.~Levis, N.~Patel, D.~Culler, and S.~Shenker, ``Trickle: {A}
  {Self}-{Regulating} {Algorithm} for {Code} {Propagation} and {Maintenance} in
  {Wireless} {Sensor} {Networks},'' in \emph{Proceedings of the 1st
  {Conference} on {Symposium} on {Networked} {Systems} {Design} and
  {Implementation} - {Volume} 1}.\hskip 1em plus 0.5em minus 0.4em\relax USENIX
  Association, 2004.

\bibitem{panta_stream_2007}
R.~Panta, I.~Khalil, and S.~Bagchi, ``Stream: {Low} {Overhead} {Wireless}
  {Reprogramming} for {Sensor} {Networks},'' in \emph{Proc. of IEEE Infocom},
  2007, pp. 928 -- 936.

\bibitem{huang_cord_2008}
L.~Huang and S.~Setia, ``{CORD}: {Energy}-{Efficient} {Reliable} {Bulk} {Data}
  {Dissemination} in {Sensor} {Networks},'' in \emph{{IEEE} {INFOCOM} 2008 -
  {The} 27th {Conference} on {Computer} {Communications}}, 2008, pp. 574--582.

\bibitem{hill_mica_2002}
J.~Hill and D.~Culler, ``Mica: a wireless platform for deeply embedded
  networks,'' \emph{IEEE Micro}, vol.~22, no.~6, pp. 12--24, 2002.

\bibitem{djamaa_optimizing_2015}
B.~Djamaa and M.~Richardson, ``Optimizing the {Trickle} {Algorithm},''
  \emph{IEEE Communications Letters}, vol.~19, no.~5, pp. 819--822, 2015.

\bibitem{ghaleb_e-trickle_2015}
B.~Ghaleb, A.~Al-Dubai, and E.~Ekonomou, ``E-{Trickle}: {Enhanced} {Trickle}
  {Algorithm} for {Low}-{Power} and {Lossy} {Networks},'' in \emph{2015 {IEEE}
  {International} {Conference} on {Computer} and {Information} {Technology};
  {Ubiquitous} {Computing} and {Communications}; {Dependable}, {Autonomic} and
  {Secure} {Computing}; {Pervasive} {Intelligence} and {Computing}}.\hskip 1em
  plus 0.5em minus 0.4em\relax IEEE, 2015, pp. 1123--1129.

\bibitem{byers_digital_1998}
J.~Byers, M.~Luby, M.~Mitzenmacher, and A.~Rege, ``A digital fountain approach
  to reliable distribution of bulk data,'' \emph{ACM SIGCOMM Computer
  Communication Review}, pp. 56--67, 1998.

\bibitem{ho_random_2006}
T.~Ho, M.~Medard, R.~Koetter, D.~R. Karger, M.~Effros, J.~Shi, and B.~Leong,
  ``A {Random} {Linear} {Network} {Coding} {Approach} to {Multicast},''
  \emph{IEEE Transactions on Information Theory}, pp. 4413--4430, 2006.

\bibitem{prabal_dutta_design_2005}
{Prabal D.}, M.~Grimmer, A.~Arora, S.~Bibyk, and D.~Culler, ``Design of a
  wireless sensor network platform for detecting rare, random, and ephemeral
  events,'' in \emph{{IPSN} 2005. {Fourth} {International} {Symposium} on
  {Information} {Processing} in {Sensor} {Networks}, 2005.}, 2005, pp.
  497--502.

\bibitem{wang_two_2005}
B.~Wang, Y.~Chen, H.~Gu, J.~Yang, and T.~Zhao, ``Two {Energy}-{Efficient},
  {Timesaving} {Improvement} {Mechanisms} of {Network} {Reprogramming} in
  {Wireless} {Sensor} {Network},'' in \emph{Embedded Software and
  Systems}.\hskip 1em plus 0.5em minus 0.4em\relax Springer Berlin Heidelberg,
  2005, pp. 473--483.

\bibitem{park_scalable_2004}
S.~Park, R.~Vedantham, R.~Sivakumar, and I.~Akyildiz, ``A scalable approach for
  reliable downstream data delivery in wireless sensor networks,'' in
  \emph{Proceedings of the 5th {ACM} international symposium on {Mobile} ad hoc
  networking and computing - {MobiHoc} '04}, 2004.

\bibitem{guha_approximation_1998}
S.~Guha and S.~Khuller, ``Approximation {Algorithms} for {Connected}
  {Dominating} {Sets},'' \emph{Algorithmica}, pp. 374--387, 1998.

\bibitem{cheng_virtual_2006}
X.~Cheng, M.~Ding, D.~Du, and X.~Jia, ``Virtual backbone construction in
  multihop ad hoc wireless networks,'' \emph{Wireless Communications and Mobile
  Computing}, pp. 183--190, 2006.

\bibitem{liu_tinyecc_2008}
A.~Liu and P.~Ning, ``{TinyECC}: {A} {Configurable} {Library} for {Elliptic}
  {Curve} {Cryptography} in {Wireless} {Sensor} {Networks},'' in \emph{2008
  {International} {Conference} on {Information} {Processing} in {Sensor}
  {Networks} (ipsn 2008)}, 2008, pp. 245--256.

\bibitem{hyun_seluge_2008}
S.~Hyun, P.~Ning, A.~Liu, and W.~Du, ``Seluge: {Secure} and {DoS}-{Resistant}
  {Code} {Dissemination} in {Wireless} {Sensor} {Networks},'' in \emph{2008
  {International} {Conference} on {Information} {Processing} in {Sensor}
  {Networks} (ipsn 2008)}, 2008, pp. 445--456.

\bibitem{shoufan_fast_2010}
A.~Shoufan and N.~Huber, ``A fast hash tree generator for {Merkle} signature
  scheme,'' in \emph{Proceedings of 2010 {IEEE} {International} {Symposium} on
  {Circuits} and {Systems}}, 2010, pp. 3945--3948.

\bibitem{ning_mitigating_2008}
P.~Ning, A.~Liu, and W.~Du, ``Mitigating {DoS} attacks against broadcast
  authentication in wireless sensor networks,'' \emph{ACM Transactions on
  Sensor Networks}, vol.~4, pp. 1--35, 2008.

\bibitem{lanigan_sluice_2006}
P.~Lanigan, R.~Gandhi, and P.~Narasimhan, ``Sluice: {Secure} {Dissemination} of
  {Code} {Updates} in {Sensor} {Networks},'' in \emph{26th {IEEE}
  {International} {Conference} on {Distributed} {Computing} {Systems}
  ({ICDCS}'06)}, 2006.

\bibitem{dutta_securing_2006}
P.~Dutta, J.~Hui, D.~Chu, and D.~Culler, ``Securing the deluge {Network}
  programming system,'' in \emph{{IPSN}}, 2006.

\bibitem{zandberg_2019}
K.~Zandberg, K.~Schleiser, F.~Acosta, H.~Tschofenig, and E.~Baccelli, ``Secure
  firmware updates for constrained iot devices using open standards: A reality
  check,'' \emph{IEEE Access}, 2019.

\bibitem{asokan_2018}
N.~Asokan, T.~Nyman, N.~Rattanavipanon, A.~Sadeghu, and G.~Tsudik, ``Assured:
  Architecture for secure software update of realistic embedded devices,''
  \emph{IEEE Transactions On Computer-aided Design Of Integrated Circuits and
  Systems}, pp. 2290--2300, 2018.

\bibitem{Eldefrawy_2017}
K.~Eldefrawy, N.~Rattanavipanon, and G.~Tsudik, ``Hydra: hybrid design for
  remote attestation (using a formally verified microkernel),'' in
  \emph{Proceedings of ACM WiSec}, 2017, pp. 99--110.

\bibitem{pinto_2019}
S.~Pinto and N.~Santos, ``Demystifying arm trustzone: A comprehensive survey,''
  \emph{ACM Computing Surveys}, 2019.

\bibitem{doddapaneni_secure_2017}
K.~Doddapaneni, R.~Lakkundi, S.~Rao, S.~G. Kulkarni, and B.~Bhat, ``Secure
  {FoTA} {Object} for {IoT},'' in \emph{2017 {IEEE} 42nd {Conference} on
  {Local} {Computer} {Networks} {Workshops} ({LCN} {Workshops})}, 2017, pp.
  154--159.

\bibitem{plat:Menderio}
\BIBentryALTinterwordspacing
``Mender - {O}pen source over-the-air software updates for {L}inux devices.''
  [Online]. Available: \url{https://mender.io/}
\BIBentrySTDinterwordspacing

\bibitem{plat:ArmPelion}
\BIBentryALTinterwordspacing
``Arm {P}elion {I}o{T} {P}latform.'' [Online]. Available:
  \url{https://www.pelion.com}
\BIBentrySTDinterwordspacing

\bibitem{plat:Balena}
\BIBentryALTinterwordspacing
``Balena - the complete {I}o{T} fleet management platform.'' [Online].
  Available: \url{https://www.balena.io}
\BIBentrySTDinterwordspacing

\bibitem{plat:Particle}
\BIBentryALTinterwordspacing
``Particle.'' [Online]. Available: \url{https://www.particle.io/}
\BIBentrySTDinterwordspacing

\bibitem{plat:FreeRTOS}
\BIBentryALTinterwordspacing
``Free{R}{T}{O}{S} - real-time operating system for microcontrollers.''
  [Online]. Available: \url{https://www.freertos.org/}
\BIBentrySTDinterwordspacing

\bibitem{plat:YoctoProject}
\BIBentryALTinterwordspacing
``Yocto {P}roject.'' [Online]. Available: \url{https://www.yoctoproject.org/}
\BIBentrySTDinterwordspacing

\bibitem{plat:AWSIoT}
\BIBentryALTinterwordspacing
``Amazon {W}eb {S}ervices {I}o{T}.'' [Online]. Available:
  \url{https://aws.amazon.com/iot}
\BIBentrySTDinterwordspacing

\end{thebibliography}

%








\end{document}